\documentstyle[prl,aps,psfig]{revtex}

\begin{document}
\draft


\title{Thermodynamics of hexagonal-close-packed iron \\
under Earth's core conditions}
\author{D. Alf\`{e}$^1$, G. D. Price$^1$ and M. J. Gillan$^2$
\smallskip \\
$^1$Research School of Geological and Geophysical Sciences \\
Birkbeck and University College London \\
Gower Street, London WC1E 6BT, UK
\smallskip \\
$^2$Physics and Astronomy Department, University College London \\
Gower Street, London WC1E 6BT, UK}
\maketitle

\begin{abstract}
The free energy and other thermodynamic properties of
hexagonal-close-packed iron are calculated by direct {\em ab initio}
methods over a wide range of pressures and temperatures
relevant to the Earth's core. The {\em ab initio} calculations
are based on density-functional theory in the generalised-gradient
approximation, and are performed using the projector augmented
wave (PAW) approach. Thermal excitation of electrons
is fully included. The Helmholtz free energy consists of
three parts, associated with the rigid perfect lattice,
harmonic lattice vibrations, and anharmonic contributions,
and the technical problems of calculating these
parts to high precision are investigated. The harmonic
part is obtained by computing the phonon frequencies
over the entire Brillouin zone, and by summation of the free-energy
contributions associated with the phonon modes. The anharmonic
part is computed by the technique of thermodynamic integration
using carefully designed reference systems. Detailed results are presented
for the pressure, specific heat, bulk modulus, expansion coefficient
and Gr\"{u}neisen parameter, and comparisons are made with
values obtained from diamond-anvil-cell
and shock experiments.
\end{abstract}

\pacs{PACS numbers: 
71.15.-m 
64.10.+h 
62.50.+p 
}

\section{Introduction}
\label{intro}

{\em Ab-initio} techniques based on density-functional theory (DFT)
have played a key role for several years in the study of matter
under extreme conditions~\cite{stixrude98}.
With recent progress in the direct
{\em ab initio} calculation of thermodynamic free
energies~\cite{sugino95,dewijs98a,alfe99a},
there is now great scope for the systematic and accurate
calculation of thermodynamic properties over a wide range
of conditions. We present here extensive
DFT calculations of the free
energy of hexagonal-close-packed (h.c.p.)
iron under Earth's core conditions, which we have used to obtain
results for a number of other thermodynamic quantities, including
the bulk modulus, expansion coefficient,
specific heat and Gr\"{u}neisen parameter. For some of these
we can make direct comparisons with experimental data, which
support the accuracy and realism of the calculations; for
others, the calculations supply information that is not
yet available from experiments. An important ambition
of the work is to determine thermodynamic functions without
making any significant statistical-mechanical or electronic-structure
approximations, other than those required by DFT itself, and
we shall argue that we come close to achieving this. The techniques
we have developed are rather general, and we believe
they will find application to many other problems concerning
matter under extreme conditions.

The importance of understanding the high-pressure and high-temperature
properties of iron can be appreciated by recalling that the Earth's
core accounts for about 30~\% of the mass of the entire
Earth, and consists mainly of iron~\cite{poirier91}.
In fact, the liquid outer
core is somewhat less dense than pure iron, and is generally
accepted to contain light impurities such as S, O, Si or
H~\cite{poirier94}; probably
the density of the solid inner core is also
significantly reduced by impurities~\cite{masters90,stixrude97}.
Nevertheless, the thermodynamic
properties of pure iron are fundamental to understanding
the more complex real material in the core, and a large
experimental effort has been devoted to measuring them. The
difficulties are severe, because the pressure range of
interest extends from 100~GPa up to nearly 400~GPa, and the
temperature goes from {\em ca.}~3000~K to
perhaps 7000~K -- the temperature at the centre of the core
is subject to an uncertainty of at least 1000~K.


Static compression experiments with the diamond anvil cell (DAC)
have been performed on Fe up to 300~GPa
at room temperature~\cite{mao90}, and DAC experiments at temperatures
as high as 3700~K have been reported up to
200~GPa~\cite{boehler90,boehler93,saxena93,saxena94,saxena95,saxena96,jephcoat96,andrault97,shen98}. 
Our present knowledge of the
phase diagram of Fe comes mainly from these experiments, though
there are still controversies. Suffice it say that for pressures
$p$ above {\em ca.}~60~GPa and temperatures $T$ below {\em ca.}~1500~K
it is generally accepted that the stable phase is hexagonal
close packed (h.c.p.). Recent DAC diffraction experiments~\cite{shen98}
indicate that h.c.p. is actually stable for all temperatures up
to the melting line for $p > 60$~GPa, but earlier work claimed
that there is another phase, usually called $\beta$, in a region below
the melting line for pressures above {\em ca.}~40~GPa. 
The existing evidence suggests
that, if the $\beta$-phase is thermodynamically stable, its structure
could either be double-h.c.p.~\cite{saxena95,saxena96} or
orthorhombic~\cite{andrault97}, and in either case is closely related
to the usual h.c.p. structure. According to very recent theoretical
work~\cite{vocadlo99}, 
h.c.p. is thermodynamically slightly more stable than double-h.c.p.
at Earth's core pressures and temperatures. The evidence for
the stability of h.c.p. over much of the high-temperature/high-pressure
phase diagram is our motivation for concentrating on this phase
in the present work.

DAC measurements have given some information about thermodynamic
quantities up to pressures of a few tens of GPa,
but beyond this shock experiments (see e.g. 
Refs.~\cite{jeanloz79,brown86,yoo93})
have no competitors.
These experiments give direct values of the pressure as a function
of volume~\cite{brown86} on the Hugoniot curve, and have also been
used to obtain information about the adiabatic bulk modulus
and some other thermodynamic quantities on this curve.
These data will be important in validating
our calculations. Temperature is difficult to measure
reliably in shock experiments~\cite{yoo93}, and we believe that our
{\em ab initio} results may be valuable in providing the
needed calibration. 

The difficulties and uncertainties of experiments have
stimulated many theoretical efforts. Some of the theoretical
work has been based on simple atomistic
models, such as a representation of the total energy as
a sum of pair potentials~\cite{matsui97},
or the more sophisticated embedded-atom
model~\cite{belonoshko97}. Such models can be useful,
but for accuracy and reliability
they cannot match high-quality {\em ab initio}
calculations based on density-functional
theory (DFT)~\cite{generaldft}. The accuracy
of DFT depends very much on the approximation used for
the electronic exchange-correlation energy $E_{\rm xc}$.
It is known that the local density approximation (LDA)
is not fully satisfactory for Fe~\cite{wang85}, but that modern
generalised-gradient approximations (GGA) reproduce a wide
range of properties very accurately. These include the
equilibrium lattice parameter, bulk modulus and magnetic
moment of body-centred cubic (b.c.c.) Fe at ambient
pressures~\cite{stixrude94,soderlind96,vocadlo97}
and the phonon dispersion relations of the b.c.c. phase~\cite{vocadlo99}.
There has been much DFT work
on different crystal structures of Fe at high pressures, and
experimental low-temperature results for the pressure as
a function of volume $p(V)$ up to $p = 300$~GPa for the
hexagonal close packed (h.c.p.) structure are
accurately predicted~\cite{mao90}. Further evidence for the accuracy
of DFT comes from the successful prediction of the b.c.c.
to h.c.p. transition
pressure~\cite{stixrude94,soderlind96}. With {\em ab-initio}
molecular dynamics, DFT calculations can also be performed
on the liquid state, and we have reported extensive calculations
both on pure liquid Fe~\cite{vocadlo97,dewijs98b,alfe99b}
and on liquid Fe/S and Fe/O 
alloys~\cite{alfe98a,alfe98b,alfe99c}.

Recently, work has been reported~\cite{stixrude97,wasserman96}
in which the thermal properties of close-packed
crystalline Fe under
Earth's core conditions are calculated using {\em ab initio} methods.
In fact, the work itself was based on a tight-binding
representation of the total energy, but this was parameterised
using extensive {\em ab initio} data. The authors did not
attempt to perform the statistical-mechanical calculations
exactly, but instead used the so-called `particle in a cell'
approximation~\cite{holt70}, in which vibrational correlations between atoms
are ignored. In spite of these limitations, the work yielded
impressive agreement with shock data. We shall make comparisons
with this work at various points in the
present paper.

The present DFT work is based on the GGA known as
Perdew-Wang 1991~\cite{wang91,perdew92}. The choice
of functional for exchange-correlation energy $E_{\rm xc}$
completely determines the free energy and all other thermodynamic
quantities. This statement is important enough to be worth elaborating.
It is clear that a given $E_{\rm xc}$ exactly determines the
total energy of the system $U ( {\bf R}_1 , \ldots {\bf R}_N )$
for given positions $\{ {\bf R}_i \}$ of the atoms. But through
the standard formulas of statistical mechanics, the function
$U ( {\bf R}_1 , \ldots {\bf R}_N )$ exactly determines
the free energy. So, provided no further electronic-stucture
approximations are made in calculating $U ( {\bf R}_1 , \ldots
{\bf R}_N )$ from $E_{\rm xc}$, and no statistical-mechanical
approximations are made in obtaining $F$ from
$U ( {\bf R}_1 , \ldots {\bf R}_N )$, then $E_{\rm xc}$
exactly determines $F$.
The calculation of $U$ from $E_{\rm xc}$
has been discussed over many years by many authors. The work
presented here is based on the projector-augmented wave (PAW)
implementation of DFT~\cite{blochl94,kresse99},
which is an all-electron technique
similar to other standard implementations such as
full-potential linear augmented plane waves (FLAPW)~\cite{wei85},
as well as being very closely related to the
ultrasoft pseudopotential (USPP) method~\cite{vanderbilt90}.
In principle,
PAW allows one to compute $U$ with any required precision for a given
$E_{\rm xc}$. In practical tests on Fe and other systems~\cite{kresse99},
the technique has been shown to yield results that are
almost indistinguishable from calculations based on FLAPW, USPP
and other DFT implementations -- provided the same $E_{\rm xc}$
is used, of course. We aim to demonstrate in this work that $F$
can also be computed from the {\em ab-initio}
$U ( {\bf R}_1 , \ldots {\bf R}_N )$ to any required
precision. In this sense, all the approximations
made in calculating thermodynamic quantities are completely
controlled, with the sole exception of $E_{\rm xc}$ itself.

To clarify the precision we are aiming for in the calculation of $F$,
we need to explain that one of the wider objectives of this work
is the {\em ab initio} determination of the high-pressure
melting properties of Fe, to be reported in detail 
elsewhere~\cite{alfe99d}.
Our approach to melting starts from the basic principle
that at coexistence the Gibbs free energies $G_{\rm sol} ( p , T )$
and $G_{\rm liq} ( p , T )$ of solid and liquid are equal.
But for a given pressure, the curves of $G_{\rm sol} ( p , T )$
and $G_{\rm liq} ( p , T )$ cross at a shallow angle. The difference
of slopes $( \partial G_{\rm sol} / \partial T )_p = S_{\rm sol}$
and $( \partial G_{\rm liq} / \partial T )_p = S_{\rm liq}$ is
equal to the entropy of fusion $S_{\rm m} \equiv S_{\rm liq} - S_{\rm sol}$,
which is comparable to $k_{\rm B}$ per atom. This means
that to get the melting temperature within an error of
$\delta T$, the non-cancelling errors in $G_{\rm sol}$
and $G_{\rm liq}$ must not exceed {\em ca.}~$k_{\rm B} \delta T$.
Ideally, we should like to calculate the melting temperature
to within {\em ca.}~100~K, so that non-cancelling
errors must be reduced to the level of {\em ca.}~10~meV.
Our original ambition for the present work on h.c.p. Fe was
to obtain $F$ from the given {\em ab-initio} $U ( {\bf R}_1 ,
\ldots {\bf R}_N )$ to this precision, and to demonstrate
that this has been achieved. As we shall see, this target has
probably not been attained, but we miss it by only small factor,
which will be estimated.

We shall present results for thermodynamic quantities for pressures
$50 < p < 400$~GPa and temperatures $2000 < T < 6000$~K. This is
a far wider range than is strictly needed for understanding
the inner core, where pressures span the range $330 < p < 364$~GPa
and $T$ is believed to be in the region of $5000 - 6000$~K. However,
the wider range is essential in making comparisons with the
available laboratory data. We set the lower limit of 2000~K
for our $T$ range because this is the lowest $T$ that
has been proposed for equilibrium between the h.c.p. crystal
and the liquid (at lower $T$, melting occurs from the f.c.c. phase).

In the next Section, we summarize the {\em ab initio}
techniques, and give a detailed explanation of the statistical-mechanical
techniques. The three Sections after that present our
investigations of the three main components of the
free energy, associated with the rigid perfect lattice,
harmonic lattice vibrations, and anharmonic contributions,
probing in each case the technical measures that must be
taken to achieve our target precision. Sec.~6 reports
our results for all the thermodynamic quantities derived
from the free energy, with comparisons wherever possible
with experimental measurements and previous theoretical values.
Overall discussion and conclusions are given in Sec.~7. The
implications of our results for deepening our understanding
of the Earth's core will be analysed elsewhere.

\section{Techniques}
\label{sec:techniques}

\subsection{Ab initio techniques}
\label{sec:abinitio}

The use of DFT to calculate the energetics of many-atom systems
has been extensively reviewed~\cite{generaldft}. However, a special feature
of the present work is that thermal electronic excitations
are crucially important, and we need to clarify the theoretical
framework in which the calculations are done.

A fundamental quantity in this work is the electronic free
energy $U ( {\bf R}_1 , \ldots {\bf R}_N ; T_{\rm el} )$ calculated
at electronic temperature $T_{\rm el}$ with the $N$ nuclei
fixed at positions ${\bf R}_1 , \ldots {\bf R}_N$. Throughout
most of the work, the statistical mechanics of the nuclei will
be treated in the classical limit. We lose nothing by doing this,
since we shall demonstrate later that quantum corrections
are completely negligible under the conditions of interest. The
Helmholtz free energy of the whole system is then:
\begin{equation}
\label{eqn:helmholtz}
F = - k_{\rm B} T  \ln \left\{ \frac{1}{N ! \Lambda^{3 N}}
\int d {\bf R}_1 \ldots d {\bf R}_N \,
\exp \left[ - \beta U ( {\bf R}_1 , \ldots {\bf R}_N ; T_{\rm el} )
\right] \right\} \; ,
\end{equation}
where $\Lambda = h / ( 2 \pi M k_{\rm B} T )^{1/2}$ is
the thermal wavelength, with $M$ being the nuclear mass, and
$\beta = 1 / k_{\rm B} T$. In practice, the electronic and
nuclear degrees of freedom are in thermal equilibrium with each other,
so that $T_{\rm el} = T$, but it will be useful to preserve the
logical distinction between the two. Although
$U ( {\bf R}_1 , \ldots {\bf R}_N ; T_{\rm el} )$ is
actually a {\em free} energy, we will generally refer to it simply
as the total-energy function, to avoid confusion with the
overall free energy $F$. It is clear from Eqn~(\ref{eqn:helmholtz}) that
in the calculation of $F$, and hence of all other thermodynamic
quantities, it makes no difference that $U$ is a free energy;
it is simply the object that plays the role of the total
energy in the statistical mechanics of the nuclei.

The DFT formulation of the electronic free energy $U$ has been
standard since the work of Mermin~\cite{mermin65}, and has frequently been
used in practical calculations~\cite{gillan89,wentzcovitch92}, 
though usually as a mere
technical device for economising on Brillouin-zone
sampling, rather than because electronic thermal excitation
was important. The essence of finite-temperature DFT is that
$U$ is given by:
\begin{equation}
U = E - T S \; ,
\end{equation}
where the DFT total energy $E$ is the usual sum of kinetic,
electron-nucleus, Hartree and exchange-correlation energy terms,
and $S$ is the electronic entropy:
\begin{equation}
S = - 2 k_{\rm B} \sum_i \left[ f_i \ln f_i + ( 1 - f_i )
\ln ( 1 - f_i ) \right] \; ,
\label{eqn:entropy}
\end{equation}
with $f_i$ the thermal (Fermi-Dirac) occupation number
of orbital $i$. The electronic kinetic energy and other parts
of $E$ also contain the occupation numbers. We point out that in
exact DFT theory the exchange-correlation (free) energy
$E_{\rm xc}$ has an explicit temperature dependence.
Very little is known about its dependence on temperature,
and we assume throughout this work that $E_{\rm xc}$ has
its zero-temperature form.

The PAW implementation of DFT has been described in detail
in previous papers~\cite{blochl94,kresse99}.
The present calculations were done
using the VASP code~\cite{kresse96a,kresse96b}.
The details of the core radii,
augmentation-charge cut-offs, etc are exactly as in our recent
PAW work on liquid Fe~\cite{alfe99b}.
Our division into valence and core
states is also the same: the 3$p$ electrons are treated as
core states, but their response to the high compression
is represented by an effective pair potential, with
the latter constructed using PAW calculations in which
the 3$p$ states are explicitly included as valence states.
Further technical details are as follows. All the calculations
are based on the form of GGA known as
Perdew-Wang 1991~\cite{wang91,perdew92}.
Brillouin-zone sampling was performed using
Monkhorst-Pack special points~\cite{monkhorst76},
and the detailed form of sampling will be noted where
appropriate. The plane-wave cut-off of 300~eV was used,
exactly as our PAW work on liquid Fe.

\subsection{Components of the free energy}
\label{sec:components}
Our {\em ab initio} calculations of thermodynamic properties are
based on a separation of the Helmholtz free energy $F$ into
the three components mentioned in the Introduction, which are
associated with the rigid perfect crystal,
harmonic lattice vibrations, and anharmonic contributions.

To explain this separation, we start from the expression
for $F$ given in Eqn~(\ref{eqn:helmholtz}).
We let $F_{\rm perf} ( T_{\rm el} ) \equiv
U ( {\bf R}_1^0 , \ldots {\bf R}_N^0 ; T_{\rm el} )$ denote
the total free energy of the system when all atoms
are fixed at their perfect-lattice positions ${\bf R}_I^0$,
and write $U ( {\bf R}_1 , \ldots {\bf R}_N ; T_{\rm el} ) =
F_{\rm perf} ( T_{\rm el} ) +
U_{\rm vib} ( {\bf R}_1 , \ldots {\bf R}_N ; T_{\rm el} )$,
which defines the vibrational energy $U_{\rm vib}$. Then
it follows from eqn~(\ref{eqn:helmholtz}) that
\begin{equation}
F =F_{\rm perf} + F_{\rm vib} \; ,
\end{equation}
where the vibrational free energy $F_{\rm vib}$ is given by:
\begin{equation}
F_{\rm vib} = - k_{\rm B} T \ln \left \{
\frac{1}{\Lambda^{3N}} \int d {\bf R}_1 \ldots d {\bf R}_N \,
\exp [ - \beta U_{\rm vib} ( {\bf R}_1 , \ldots {\bf R}_N ; T_{\rm el} ) ]
\right\} \; .
\end{equation}
(Note that we now omit the factor $N!$ from the partition
function, since every atom is assumed to be confined to
its own lattice site.)
The vibrational energy $U_{\rm vib}$ can be further separated
into harmonic and anharmonic parts ($U_{\rm vib} = U_{\rm harm} +
U_{\rm anharm}$), in terms of which we can define the harmonic
vibrational free energy $F_{\rm harm}$:
\begin{equation}
F_{\rm harm} = - k_{\rm B} T \ln \left\{
\frac{1}{\Lambda^{3N}} \int d {\bf R}_1 \ldots d {\bf R}_N \,
\exp [ - \beta U_{\rm harm} ( {\bf R}_1 , \ldots {\bf R}_N ; T_{\rm el} ) ]
\right\} \; ,
\label{eqn:fharm0}
\end{equation}
with the anharmonic free energy being the remainder $F_{\rm anharm} =
F_{\rm vib} - F_{\rm harm}$. The harmonic energy $U_{\rm harm}$
is defined in the obvious way:
\begin{equation}
U_{\rm harm} = \frac{1}{2} \sum_{I , J}
{\bf u}_I \cdot \left( \nabla_I \nabla_J U \right) \cdot {\bf u}_J \; ,
\end{equation}
where ${\bf u}_I$ is the displacement of atom $I$ from its
perfect-lattice position (${\bf u}_I \equiv {\bf R}_I - {\bf R}_I^0$)
and the double gradient of the {\em ab initio} total energy is evaluated
with all atoms at their perfect-lattice
positions. Since we are dealing with a crystal, we shall
usually prefer to rewrite $U_{\rm harm}$ in the
more explicit form:
\begin{equation}
U_{\rm harm} = \frac{1}{2} \sum_{l s \alpha , l^\prime t \beta}
u_{l s \alpha} \Phi_{l s \alpha , l^\prime t \beta}
u_{l^\prime t \beta} \; ,
\end{equation}
where $u_{l s \alpha}$ is the $\alpha$th Cartesian component
of the displacement of atom number $s$ in primitive
cell number $l$, and $\Phi_{l s \alpha , l^\prime t \beta}$ is
the force-constant matrix.
It should be noted that the present separation
of $F$ does not represent a separation into electronic
and nuclear contributions, since thermal electronic excitations
influence all three parts of $F$.

Since all other thermodynamic functions can be obtained by
taking appropriate derivatives of the Helmholtz free
energy, the separation of $F$ into components implies
a similar separation of other quantities. For example,
the pressure $p = - ( \partial F / \partial V )_T$ is $p_{\rm perf} +
p_{\rm harm} + p_{\rm anharm}$, where $p_{\rm perf} =
- ( \partial F_{\rm perf} / \partial V )_T$, and similarly for the
components $p_{\rm harm}$ and $p_{\rm anharm}$.

\subsection{Phonon frequencies}
\label{sec:phonons}

The free energy of a harmonic oscillator of frequency
$\omega$ is $k_{\rm B} T \ln \left(
\exp ( \frac{1}{2} \beta \hbar \omega ) -
\exp ( - \frac{1}{2} \beta \hbar \omega ) \right)$, which
has the high-temperature expansion $k_{\rm B} T \ln ( \beta \hbar \omega ) +
k_{\rm B} T \left[ \frac{1}{24} ( \beta \hbar \omega )^2 +
O ( ( \beta \hbar \omega )^4 ) \right]$, so that the harmonic
free energy per atom of the vibrating crystal in the classical limit is:
\begin{equation}
F_{\rm harm} = \frac{3 k_{\rm B} T}{N_{{\rm k} s}} \sum_{{\bf k} s}
\ln ( \beta \hbar \omega_{{\bf k} s} ) \; ,
\label{eqn:fharm}
\end{equation}
where $\omega_{{\bf k} s}$ is the frequency of phonon branch
$s$ at wavevector ${\bf k}$ and the sum goes over the first
Brillouin zone, with $N_{{\bf k} s}$ the total number
of $k$-points and branches in the sum. It will sometimes be useful
to express this in terms of the geometric average
$\bar{\omega}$ of the phonon frequencies, defined as:
\begin{equation}
\ln \bar{\omega} = \frac{1}{N_{{\bf k} s}} \sum_{{\bf k} s}
\ln ( \omega_{{\bf k} s} ) \; ,
\end{equation}
which allows us to write:
\begin{equation}
F_{\rm harm} = 3 k_{\rm B} T \ln ( \beta \hbar \bar{\omega} ) \; .
\end{equation}

The central quantity in the calculation of the frequencies
is the force-constant matrix $\Phi_{l s \alpha , l^\prime t \beta}$,
since the frequencies at wavevector ${\bf k}$ are
the eigenvalues of the dynamical matrix $D_{s \alpha , t \beta}$,
defined as:
\begin{equation}
D_{s \alpha , t \beta} ( {\bf k} ) =
\frac{1}{M} \sum_{l^\prime}
\Phi_{l s \alpha , l^\prime t \beta}
\exp \left[ i {\bf k} \cdot ( {\bf R}_{l^\prime t}^0 -
{\bf R}_{l s}^0 ) \right] \; ,
\end{equation}
where ${\bf R}_{l s}^0$ is the perfect-lattice
position of atom $s$ in primitive cell number $l$.
If we have the complete force-constant matrix, then
$D_{s \alpha , t \beta}$ and hence the frequencies
$\omega_{{\rm k} s}$ can be obtained at any ${\bf k}$,
so that $\bar{\omega}$ can be computed to any required precision.
In principle, the elements of $\Phi_{l s \alpha , l^\prime t \beta}$
are non-zero for arbitrarily large separations
$\mid {\bf R}_{l^\prime t}^0 -
{\bf R}_{l s}^0 \mid$, but in practice they decay rapidly
with separation, so that a key issue in
achieving our target precision is the cut-off distance beyond
which the elements can be neglected.

We calculate $\Phi_{l s \alpha , l^\prime t \beta}$ by the small-displacement
method, in a way similar to that described in
Ref.~\cite{kresse95}. The basic
principle is that $\Phi_{l s \alpha , l^\prime t \beta}$ describes
the proportionality between displacements and forces. If the atoms
are given small displacements $u_{l s \alpha}$ from their
perfect-lattice positions, then to linear order
the forces $F_{l s \alpha}$ are:
\begin{equation}
F_{l s \alpha} = - \sum_{l^\prime t \beta}
\Phi_{l s \alpha , l^\prime t \beta} u_{l^\prime t \beta} \; .
\end{equation}
Within the {\em ab initio} scheme, all the elements
$\Phi_{l s \alpha , l^\prime t \beta}$ are obtained for
a given $l^\prime t \beta$ by introducing a small
displacement $u_{l^\prime t \beta}$, all other displacements
being zero, minimising the electronic free energy, and
evaluating all the forces $F_{l s \alpha}$. In practice,
the displacement amplitude $u_{l^\prime t \beta}$ must be made
small enough to ensure linearity to the required precision, and this
sets the precision with which the electronic free energy must
be minimised.

By translational symmetry, the entire force-constant matrix is obtained
by making three independent displacements for each atom in the
primitive cell, and this means that no more than $3 N_{\rm bas}$
calculations are needed, where $N_{\rm bas}$ is the number of atoms
in the primitive cell. This number can be reduced
by symmetry. If, as in the h.c.p. crystal, all atoms
in the primitive cell are equivalent under operations of
the space group, then the entire force-constant matrix
can be obtained by making at most three displacements of a single
atom in the primitive cell: from $\Phi_{l s \alpha , l^\prime t \beta}$
for one chosen atom $l^\prime t$, one obtains
$\Phi_{l s \alpha , l^\prime t \beta}$ for all other
$l^\prime t$. Point-group symmetry reduces the number still
further if linearly independent displacements of the chosen atom
are equivalent by symmetry. This is the case in the h.c.p. structure,
since displacements in the basal plane related by rotations
about the $c$-axis by $\pm 120^\circ$ are equivalent
by symmetry; this means that two calculations, one with the
displacement along the $c$-axis, and other with the displacement
in the basal plane, suffice to obtain the entire
$\Phi_{l s \alpha , l^\prime t \beta}$ matrix. The basal-plane displacement
should be made along a symmetry direction, because the symmetry
makes the calculations more efficient. Since the exact
$\Phi_{l s \alpha , l^\prime t \beta}$ matrix  has point-group
symmetries, the calculated $\Phi_{l s \alpha , l^\prime t \beta}$
must be symmetrised to ensure that these symmetries
are respected. The symmetrisation also serves to eliminate the
lowest-order non-linearities in the relation between
forces and displacements~\cite{kresse95}.

It is important to appreciate that the
$\Phi_{l s \alpha , l^\prime t \beta}$ in the formula for
$D_{s \alpha , t \beta} ( {\bf k} )$ is the force-constant
matrix in the infinite lattice, with no restriction on
the wavevector ${\bf k}$, whereas the {\em ab initio}
calculations of $\Phi_{l s \alpha , l^\prime t \beta}$
can only be done in supercell geometry. Without a further
assumption, it is strictly impossible to extract the infinite-lattice
$\Phi_{l s \alpha , l^\prime t \beta}$ from supercell
calculations, since the latter deliver information
only at wavevectors that are reciprocal lattice vectors
of the superlattice. The further assumption needed is that
the infinite-lattice $\Phi_{l s \alpha , l^\prime t \beta}$
vanishes when the separation ${\bf R}_{l^\prime t} -
{\bf R}_{l s}$ is such that the positions ${\bf R}_{l s}$
and ${\bf R}_{l^\prime t}$ lie in different
Wigner-Seitz (WS) cells of the chosen superlattice. More
precisely, if we take the WS cell centred on ${\bf R}_{l^\prime t}$,
then the infinite-lattice value of $\Phi_{l s \alpha , l^\prime t \beta}$
vanishes if ${\bf R}_{l s}$ is in a different
WS cell; it is equal to the supercell value if
${\bf R}_{l s}$ is wholly within the same WS
cell; and it is equal to the supercell value divided by
an integer $P$ if ${\bf R}_{l s}$ lies on the boundary
of the same WS cell, where $P$ is the number of WS cells having
${\bf R}_{l s}$ on their boundary. With this assumption,
the $\Phi_{l s \alpha , l^\prime t \beta}$ elements will
converge to the correct infinite-lattice
values as the dimensions of the supercell are systematically
increased.

\subsection{Anharmonicity}
\label{sec:anharmonicity}

\subsubsection{Thermodynamic integration}
\label{sec:thermint}

Although we shall show that the anharmonic free
energy $F_{\rm anharm}$ is numerically
fairly small, it is far more challenging to calculate than $F_{\rm perf}$
or $F_{\rm harm}$, because there is no simple formula like
eqn~(\ref{eqn:fharm}), and the direct computation of the multi-dimensional
integrals in the free-energy formulas
such as eqn~(\ref{eqn:fharm0}) is impossible.
Instead, we use the technique of thermodynamic
integration (see e.g. Ref.~\cite{frenkel96})
to obtain the difference $F_{\rm vib} - F_{\rm harm}$, as
developed in earlier papers~\cite{sugino95,dewijs98a,alfe99a}.

Thermodynamic integration is a completely general technique
for determining the difference of free energies $F_1 - F_0$
for two systems whose total-energy functions are $U_1$ and $U_0$.
The basic idea is that $F_1 - F_0$ represents the
reversible work done on
continuously and isothermally switching the energy function
from $U_0$ to $U_1$. To do this switching, a continuously
variable energy function $U_\lambda$ is defined as:
\begin{equation}
U_\lambda = ( 1 - \lambda ) U_0 + \lambda U_1 \; ,
\end{equation}
so that the energy goes from $U_0$ to $U_1$ as $\lambda$
goes from 0 to 1. In classical statistical mechanics, the work
done in an infinitesimal change $d \lambda$ is:
\begin{equation}
d F = \langle d U_\lambda / d \lambda \rangle_\lambda d \lambda =
\langle U_1 - U_0 \rangle_\lambda d \lambda \; ,
\end{equation}
where $\langle \, \cdot \, \rangle_\lambda$ represents the
thermal average evaluated for the system governed by $U_\lambda$.
It follows that:
\begin{equation}
F_1 - F_0 = \int_0^1 d \lambda \, \langle U_1 - U_0 \rangle_\lambda \, .
\end{equation}
In practice, this formula can be applied by calculating
$\langle U_1 - U_0 \rangle_\lambda$ for
a suitable set of $\lambda$ values and performing the integration
numerically. The average
$\langle U_1 - U_0 \rangle_\lambda$ is evaluated by sampling
over configuration space.

For the anharmonic free energy, a possible approach is
to choose $U_0$ as $U_{\rm harm}$ and $U_1$ as $U_{\rm vib}$, so that
$F_1 - F_0$ is the anharmonic free energy $F_{\rm anharm}$. This was
the procedure used in our earlier {\em ab initio} work on the
melting of Al~\cite{dewijs98a}, and a
related technique was used by Sugino and
Car~\cite{sugino95} in
their work on Si melting. However, the calculations
are rather heavy, and the need for extensive sampling over the
electronic Brillouin zone in the {\em ab initio} calculations makes it
difficult to achieve high precision. We have now developed
a more efficient two-step procedure, in which we go first from the
harmonic {\em ab initio} system $U_{\rm harm}$ to an intermediate
reference system $U_{\rm ref}$ which closely mimics
the full {\em ab initio} total energy $U_{\rm vib}$; in the
second step, we go from $U_{\rm ref}$ to
$U_{\rm vib}$. The anharmonic free energy is thus represented as:
\begin{equation}
F_{\rm anharm} = ( F_{\rm vib} - F_{\rm ref} ) +
( F_{\rm ref} - F_{\rm harm} ) \; ,
\end{equation}
and the two differences are calculated by separate thermodynamic
integrations:
\begin{eqnarray}
F_{\rm vib} - F_{\rm ref} & = & \int_0^1 d \lambda \,
\langle U_{\rm vib} - U_{\rm ref} \rangle_\lambda^{\rm vr} \nonumber \\
F_{\rm ref} - F_{\rm harm} & = & \int_0^1 d \lambda \,
\langle U_{\rm ref} - U_{\rm harm} \rangle_\lambda^{\rm rh} \; .
\end{eqnarray}
To distinguish clearly between these two parts of the calculation,
we denote by $\langle \, \cdot \, \rangle_\lambda^{\rm rh}$ the
thermal average taken in the ensemble generated by the
switched total energy $U_\lambda^{\rm rh} \equiv
( 1 - \lambda ) U_{\rm harm} + \lambda U_{\rm ref}$ and
by $\langle \, \cdot \, \rangle_\lambda^{\rm vr}$
the corresponding average for $U_\lambda^{\rm vr} \equiv
( 1 - \lambda ) U_{\rm ref} + \lambda U_{\rm vib}$.

The crucial point of this is that $U_{\rm ref}$ is required to
consist of an empirical model potential which quite accurately
represents both the harmonic and anharmonic parts of
the {\em ab initio} total energy $U_{\rm vib}$. Since
it is a model potential, the thermodynamic integration
for $F_{\rm ref} - F_{\rm harm}$ can be performed
with high precision on large systems. The difference
$F_{\rm vib} - F_{\rm ref}$, by contrast, involves
heavy {\em ab initio} calculations, but provided a good
$U_{\rm ref}$ can be found these are manageable. We return
to the problem of searching for a good $U_{\rm ref}$ in
Sec.~\ref{sec:reference}.

\subsubsection{Calculation of thermal averages}
\label{sec:thermav}

The calculation of thermal averages is just the standard
problem of computational statistical mechanics, and can
be accomplished by any method that allows us to draw
unbiased samples of configurations from the appropriate ensemble.
In this work, we employ molecular dynamics simulation. This
means, for example, that to calculate
$\langle U_{\rm ref} - U_{\rm harm} \rangle_\lambda^{\rm rh}$
we generate a trajectory of the system using equations
of motion derived from the total energy function
$U_\lambda^{\rm rh}$. In the usual way, an initial part of
the trajectory is discarded for equilibration, and the remainder
is used to estimate the average. The duration of
this remainder must suffice to deliver enough {\em independent}
samples to achieve the required statistical precision.

The key technical problem in calculating thermal averages in
nearly harmonic systems is that of ergodicity. In the
dynamical evolution of a perfectly harmonic system,
energy is never shared between different vibrational modes,
so that a system starting at any point in phase space fails to explore
the whole of phase space. This means that in a nearly harmonic
system exploration will be very slow and inefficient,
and it is difficult to generate statistically
independent samples. We solve this following
Ref.~\cite{dewijs98a}: the statistical sampling is performed using
Andersen molecular dynamics~\cite{andersen80},
in which the atomic velocities
are periodically randomised by drawing them from
a Maxwellian distribution. This type of simulation generates
the canonical ensemble and completely overcomes the ergodicity
problem.

\subsubsection{Reference system}
\label{sec:reference}

The computational effort needed to calculate $F_{\rm vib} - F_{\rm ref}$
is greatly reduced if the difference of total energies
$U_{\rm vib} - U_{\rm ref}$ is small, for
two reasons. First, the amount of sampling needed to calculate
$\langle U_{\rm vib} - U_{\rm ref} \rangle_\lambda$ to a given
precision is reduced if the fluctuations of $U_{\rm vib} - U_{\rm ref}$
are small; second, the variation of $\langle U_{\rm vib} -
U_{\rm ref} \rangle_\lambda$ as $\lambda$ goes from 0 to 1 is reduced.
In fact, if the fluctuations are small enough, this variation can be
neglected, and it is accurate enough to approximate
$F_{\rm vib} - F_{\rm ref} \simeq \langle U_{\rm vib} -
U_{\rm ref} \rangle_{\rm ref}$, with the average taken in the
reference ensemble. If this is not good enough, the next
approximation is readily shown to be:
\begin{equation}
F_{\rm vib} - F_{\rm ref} \simeq \langle U_{\rm vib} -
U_{\rm ref} \rangle_{\rm ref} - \frac{1}{2 k_{\rm B} T}
\left\langle \left[ U_{\rm vib} - U_{\rm ref} -
\langle U_{\rm vib} - U_{\rm ref} \rangle_{\rm ref} \rangle
\right]^2 \right\rangle_{\rm ref} \; .
\label{eqn:secondorder}
\end{equation}
Our task is therefore to search for a model $U_{\rm ref}$
for which the fluctuations of $U_{\rm vib} - U_{\rm ref}$
are as small as possible.

The question
of reference systems for Fe has already been discussed in our
recent {\em ab initio} simulation work on the
high-pressure liquid~\cite{alfe99b}.
We showed there that a remarkably good reference
model is provided by a system interacting through inverse-power
pair potentials:
\begin{equation}
U_{\rm IP} = \frac{1}{2} \sum_{I \ne J} \phi ( \mid {\bf R}_I -
{\bf R}_J \mid ) \; ,
\label{eqn:uip}
\end{equation}
where $\phi (r) = B / r^\alpha$, with $B$ and $\alpha$ adjusted
to minimise the fluctuations of the difference between
$U_{\rm IP}$ and the {\em ab initio} energy.
Unfortunately, we shall show that this is an
unsatisfactory reference model for
the solid, because the harmonic phonon dispersion relations produced by
$U_{\rm IP}$ differ markedly from the {\em ab initio} ones.
It is a particularly poor reference
model at low temperatures where anharmonic corrections are small,
because in that r\'{e}gime a good reference system must closely
resemble $U_{\rm harm}$. However, we find that $U_{\rm IP}$ becomes
an increasingly good reference system as $T$ approaches the
melting temperature. 
We therefore adopt as a general
form for the reference system a linear combination of
$U_{\rm harm}$ and $U_{\rm IP}$:
\begin{equation}
U_{\rm ref} = c_1 U_{\rm harm} + c_2 U_{\rm IP} \; .
\label{eqn:optcoef}
\end{equation}
The coefficients $c_1$ and $c_2$ are adjusted to
minimise the intensity of the fluctuations
of $U_{\rm vib} - U_{\rm ref}$ for each thermodynamic state.

Now consider in more detail how this optimisation of $U_{\rm ref}$
is to be done. In principle, the ensemble in which we have to
sample the fluctuations of $U_{\rm vib} - U_{\rm ref}$
is the one generated by the continuously switched
total energy $( 1 - \lambda ) U_{\rm ref} + \lambda U_{\rm vib}$
that governs the thermodynamic integration from
$U_{\rm ref}$ to $U_{\rm vib}$. In practice, this is essentially
the same as sampling in either of the ensembles associated
with $U_{\rm ref}$ or $U_{\rm vib}$, provided the fluctuations
of $U_{\rm vib} - U_{\rm ref}$ are indeed small. But even this
poses a problem. We are reluctant to sample in the ensemble
of $U_{\rm vib}$, because extensive (and expensive)
{\em ab initio} calculations are needed to achieve adequate
statistical accuracy. On the other hand, we cannot sample
in the ensemble of $U_{\rm ref}$ without knowing $U_{\rm ref}$,
which is what we are trying to find. We resolve this
problem by constructing an initial optimised $U_{\rm ref}$
by minimising the fluctuations in the ensemble of $U_{\rm harm}$.
We then use this initial $U_{\rm ref}$ to generate a new set of samples,
which is then used to reoptimise $U_{\rm ref}$. In principle,
we should probably repeat this procedure until $U_{\rm ref}$
ceases to vary, but in practice we stop after the second
iteration. Note that even this approach requires fully
converged {\em ab initio} calculations for a large set of
configurations. But since the configurations
are generated with the potential model $U_{\rm ref}$, statistically
independent samples are generated with much less effort than if
we were using $U_{\rm vib}$ to generate them.

\section{The rigid perfect lattice}
\label{sec:perfect}

The energy and pressure as functions of volume for h.c.p. Fe
at low temperatures (i.e temperatures at which lattice vibrations
and electronic excitations can be neglected) have been
extensively investigated both by DAC experiments~\cite{mao90} and by DFT
studies~\cite{stixrude94,soderlind96}, including
our own earlier USPP~\cite{vocadlo97} and PAW~\cite{alfe99b}
calculations. The various DFT calculations agree very closely with each
other, and reproduce the experimental $p(V)$ relation very accurately,
especially at high pressures: the difference between our PAW
pressures~\cite{alfe99b}
and the experimental values ranges from 4.5~\% at 100~GPa
to 2.5~\% at 300~GPa, these deviations being only slightly
greater than the scatter on the experimental values.

The present DFT calculations on the rigid perfect lattice
give $F_{\rm perf}$ for any chosen volume and electronic
temperature $T_{\rm el}$. In order to achieve our target
precision of 10~meV/atom for the free energy, careful
attention must be paid to electronic Brillouin-zone sampling.
All the calculations presented in this Section employ
the $15 \times 15 \times 9$ Monkhorst-Pack set, which gives 135 $k$-points
in the irreducible wedge of the Brillouin zone. The $k$-point
sampling errors with this set have been assessed by repeating selected
calculations with 520 $k$-points in the irreducible wedge.
Tests at the atomic volume $V = 8.67$~\AA$^3$ show
that the errors are 4.0, 2.0 and 0.5~meV/atom
at temperatures of 500, 1000 and 2000 respectively, so that,         
as expected, the errors decrease rapidly with increasing
$T_{\rm el}$. Since temperatures below 2000~K are not of interest here,
we conclude that with our chosen Monkhorst-Pack set
$k$-point errors are
completely negligible.

We have done direct DFT calculations at a closely spaced set
of atomic volumes going from 6.2 to 11.4~\AA$^3$ at intervals of
0.2~\AA$^3$, and at each of these volumes we have made
calculations at $T_{\rm el}$ values going from
200 to 10,000~K at intervals of 200~K. At every one of
these state points, we obtain the value of $F_{\rm perf}$.
The calculations also deliver directly the internal
energy $E_{\rm perf}$ and the entropy $S_{\rm perf}$ (see
Eqn~\ref{eqn:entropy}). The specific heat is then obtained
either by numerical differentiation of the $E_{\rm perf}$ results
($C_{\rm perf} = ( \partial E_{\rm perf} / \partial T )_V$)
or analytically from the formula for $S_{\rm perf}$
($C_{\rm perf} = T ( \partial S_{\rm perf} / \partial T )_V$). If we
ignore the temperature dependence of the Kohn-Sham energies,
$\partial S_{\rm perf} / \partial T$ can be evaluated
analytically using the text-book formula for the
Fermi-Dirac occupations numbers: $f_i =
1 / [ \exp ( \beta ( \epsilon_i - \mu )) + 1 ]$, where
$\epsilon_i$ is the Kohn-sham energy of orbital $i$ and $\mu$
is the electronic chemical potential. It was pointed
out by Wasserman {\em et al.}~\cite{wasserman96} that the
neglect of the dependence of $\epsilon_i$ on $T_{\rm tel}$
is a very accurate approximation, and out tests confirm that the errors
incurred by doing this are negligible.

Our analytically calculated results for $C_{\rm perf}$
are reported in Fig.~\ref{fig:cv_electronic} for the
temperature range $0 - 6000$~K at the atomic volumes $V = 7.0$,
8.0, 9.0 and 10.0~\AA$^3$/atom. The key point to note
is that $C_{\rm perf}$ becomes {\em large} at high temperatures,
its value of {\em ca.}~$2 k_{\rm B}$ at 6000~K being comparable
with the Dulong-Petit specific heat of lattice
vibrations ($3 k_{\rm B}$). This point about the large
magnitude of the electronic specific heat has been
emphasised in several previous
papers~\cite{stixrude97,wasserman96,boness90}, and the values reported
here are close to those calculated by
Wasserman {\em et al.}~\cite{wasserman96},
though the latter actually refer to f.c.c. Fe. This means that
full inclusion of thermal electronic excitations, as done here,
is crucial to a correct description of the thermodynamics
of Fe at core conditions.

The linear dependence of $C_{\rm perf}$ on $T$ evident
in Fig.~\ref{fig:cv_electronic}
at low $T$ ($C_{\rm perf} = \gamma T + O ( T^2 )$)
is what we expect from the standard Sommerfeld
expansion~\cite{ashcroft76} for electronic specific heat in powers of $T$,
which shows that the low-temperature slope is given
by $\gamma = \frac{1}{3} \pi^2 k_{\rm B}^2 g ( E_{\rm F} )$,
where $g ( E_{\rm F} )$ is the electronic density of states
(DOS, i.e. the number of states per unit energy per atom) at the
Fermi energy $E_{\rm F}$. Our calculated DOS at the atomic volumes
$V = 7.0$ and 10.0~\AA$^3$ (Fig.~\ref{fig:dos})
shows, as expected, that the width of the 
electronic $d$-band increases on compression, so that
$g ( E_{\rm F} )$ and hence $\gamma$ decrease with
decreasing atomic volume. As a cross-check,
we have calculated $\gamma$ directly from the density of states,
and we recover almost exactly the low temperature slope
of $C_{\rm perf}$.

In order to obtain other thermodynamic functions, we need a
fit to our $F_{\rm perf}$ results. At each temperature,
we fit the results to the standard Birch-Murnaghan
form, using exactly the procedure followed in our recent
work on the Fe/O system~\cite{alfe99c}. This involves fitting the 22 values
of $F_{\rm perf}$ at a given temperature using four fitting
parameters ($E_0$, $V_0$, $K$ and $K^\prime$ in the notation
of Ref.~\cite{alfe99c}). We find that at all temperatures the r.m.s.
fitting errors are less than 1~meV at all points. The
temperature variation of the fitting parameters is then
represented using a polynomial of sixth degree.

Electronic excitations have a significant effect on the
pressure, as can be seen by examining the $T$-dependence
of the perfect-lattice pressure $p_{\rm perf} =
- ( \partial F_{\rm perf} / \partial V )_T$. We display in
Fig.~\ref{fig:p_electronic} the thermal part
$\Delta p_{\rm perf}$ of $p_{\rm perf}$, i.e.
the difference between $p_{\rm perf}$ at a given $T$ and its
zero-temperature value. The thermal excitation of electrons
produces a positive pressure. This is what intuition would suggest,
but it is worth noting the reason. Since $F_{\rm perf} ( T_{\rm el} ) =
F_{\rm perf} (0) - \frac{1}{2} T^2 \gamma (V)$ at
low temperatures, the change of pressure due to
electronic excitations is $\Delta p = \frac{1}{2}
T^2 d \gamma / d V$ in this region. But
$d \gamma / d V > 0$, so that the electronic thermal pressure must
be positive. To put the magnitude of this pressure in context,
we recall that at the Earth's inner-core boundary (ICB) the
pressure is 330~GPa and the temperature is believed to
be in the range $5000 - 6000$~K, the atomic volume of Fe
under these conditions being {\em ca.}~7~\AA$^3$. Our results
then imply that electronic thermal pressure accounts for
{\em ca.}~4~\% of the total pressure, which is small
but significant.

\section{The harmonic crystal}
\label{sec:harmonic}

\subsection{Convergence tests}
\label{sec:convergence}
We have undertaken extensive tests to show that our target
precision of 10~meV/atom can be attained for the harmonic free
energy $F_{\rm harm}$. It is useful to note that at
$T = 6000$~K an error of 10~meV represents 2~\% of $k_{\rm B} T$,
and since $F_{\rm harm}$ is given in terms of the geometric
mean frequency $\bar{\omega}$ by $3 k_{\rm B} T \ln ( \hbar
\bar{\omega} / k_{\rm B} T )$,
we must achieve a precision of 0.7~\% in $\bar{\omega}$.
A {\em sufficient} condition for this is that we obtain the phonon
frequencies $\omega_s ( {\bf k} )$ for
all wavevectors ${\bf k}$ and branches $s$ to this
precision, but this may not be {\em necessary}, since
there can be cancellation of errors at different {\bf k}
and/or $s$. Convergence of $\bar{\omega}$ must be
ensured with respect to four main parameters: the atom
displacement used to calculate the force-constant matrix
$\Phi_{l s \alpha , l^\prime t \beta}$; the electronic
$k$-point sampling; the size of repeating cell used to obtain
$\Phi_{l s \alpha , l^\prime t \beta}$; and the density of
$k$-point mesh used in calculating $\bar{\omega}$ from
the $\omega_{{\bf k} s}$ by integration over the phonon
Brillouin zone (see Eqn~(\ref{eqn:fharm})). Convergence can be tested
separately with respect to these four parameters.

Integration over the phonon Brillouin zone is performed using
Monkhorst-Pack $k$-points~\cite{monkhorst76}. We find that the set having
364 $k$-points in the irreducible wedge achieves
a precision in $F_{\rm harm}$ of better than 1~meV/atom at all
the temperatures of interest, and this $k$-point
set was used in the calculations presented here. The effect of
atomic displacement amplitude was tested with the
force-constant matrix generated using a $2 \times 2 \times 2$
repeating cell at the atomic volume 8.67~\AA$^3$, and amplitudes
ranging from 0.0148 to 0.118~\AA\ were used. The
systematic variation of the resulting $F_{\rm harm}$ showed that
with an amplitude of 0.0148~\AA\ the error is less than 1~meV/atom.

The errors associated with electronic $k$-point sampling
in the calculation of the force-constant matrix were initially
assessed with $\Phi_{l s \alpha , l^\prime t \beta}$ obtained
from the $2 \times 2 \times 2$ repeating cell at the atomic volume
$V = 8.67$~\AA$^3$. We found that at $T_{\rm el} = 4300$~K
convergence of $F_{\rm harm}$ is obtained within 2~meV/atom if the
$3 \times 3 \times 2$ Monkhorst-Pack set of electronic
$k$-points is used. This is close to being
satisfactory, but starts to produce significant
errors at lower temperatures. Our definitive calculations
were actually done with the more extensive
$5 \times 5 \times 5$ Monkhorst-Pack electronic set. With this set,
our tests performed with $\Phi_{l s \alpha , l^\prime t \beta}$ obtained
from the $3 \times 3 \times 2$ repeating cell and $V = 9.17$~\AA$^3$
show that the electronic $k$-point error is now {\em ca.}~0.1~meV/atom
even at $T_{\rm el} = 1500$~K. At higher electronic
temperatures and with larger repeating cells, the error will of course
be even smaller.

Finally, we have tested the convergence of $F_{\rm harm}$
with respect to the size of repeating cell used to
generate $\Phi_{l s \alpha , l^\prime t \beta}$. A wide range of
different cell sizes and shapes were studied, including
$2 \times 2 \times 2$,
$3 \times 3 \times 2$, $4 \times 4 \times 4$ and $5 \times 5 \times 3$,
the largest of these containing 150 atoms in the repeating cell.
The tests showed that with the repeating cell $3 \times 3 \times 2$
the error in $F_{\rm harm}$ calculated at $V = 8.67$~\AA$^3$
at 4300~K is a little over 2~meV/atom, and we adopted
this cell size for all our calculations of
$\Phi_{l s \alpha , l^\prime t \beta}$. We expect the error to
be similar at other volumes, and to be roughly proportional to
temperature, so that it should be insignificant over the whole
range of states of interest.

\subsection{Dispersion relations, average frequency, free energy}
\label{sec:dispersion}

In Fig.~\ref{fig:phonons} we present the harmonic phonon dispersion
relations at the two atomic volumes 8.67 and 6.97~\AA$^3$
calculated with $T_{\rm el} = 4000$~K.
We are not aware of previous direct {\em ab initio}
calculations of the phonon frequencies of high-pressure h.c.p. Fe,
but there are published dispersion relations derived from
a `generalised pseudopotential' parameterisation of FP-LMTO
calculations performed by
S\"{o}derlind {\em et al.}~\cite{soderlind96} using
the LDA at the atomic volume 6.82~\AA$^3$. The agreement of
their phonon frequencies with ours is far from perfect. For
example, we find that the maximum frequency in the Brillouin
zone calculated at $V = 6.82$~\AA$^3$ is at the $\Gamma$-point
and is 21.2~THz, whereas they find the maximum frequency at
the $M$-point with the value 17.2~THz. This is not
unexpected, since they report
that the generalised pseudopotential scheme fails to reproduce
accurately some phonon frequencies calculated directly
with FP-LMTO in the f.c.c. Fe crystal~\cite{soderlind96}; in addition, the
LDA used by them is known to underestimate phonon frequencies
in Fe~\cite{vocadlo97}.

Casual inspection suggests that our dispersion curves at the two
atomic volumes are almost identical apart from an
overall scale factor. This suggestion can be judged from the
right-hand panel of Fig.~\ref{fig:phonons}, where we plot as dashed curves
the dispersion curves at $V = 8.67$~\AA$^3$ scaled by
the factor 1.409 -- the reason for choosing this factor will be
explained below. The comparison shows that the curves at the two volumes
are indeed related by a single scaling factor to within
{\em ca.}~5~\%. We also take the opportunity here
to check how well the inverse-power potential
model $U_{\rm IP}$ (see Eqn~(\ref{eqn:uip}))
reproduces phonon frequencies. To
do this, we take exactly the same parameters $B$ and $\alpha$
specifying $\phi (r)$ that reproduced well the properties
of the liquid~\cite{alfe99b}, namely $\alpha = 5.86$ and $B$ such
that for $r = 2.0$~\AA\ $\phi (r) = 1.95$~eV. The phonons
calculated from this model are compared with the
{\em ab initio} phonons at atomic volume $V = 8.67$~\AA\
in the left panel of Fig.~\ref{fig:phonons}. Although
the general form of the dispersion curves is correctly
reproduced, it is clear that the model gives only a very
rough description, with discrepancies of as much as 30~\%
for some frequencies.

We performed direct {\em ab initio} calculations of the
dispersion relations and hence the geometric mean
frequency $\bar{\omega}$ for seven volumes spaced roughly
equally from 9.72 to 6.39~\AA$^3$, and for each of these
volumes for $T_{\rm el}$ from 1000 to 10,000~K at intervals
of 500~K. The results for $\bar{\omega}$ as function of
volume are reported in Fig.~\ref{fig:omega} for the three temperatures
$T_{\em el} = 2000$, 4000 and 6000~K. We use a (natural) log-log
plot to display the results, so that the negative
slope $\gamma_{\rm ph} \equiv - d \ln \bar{\omega} / d \ln V$
is the so-called phonon Gr\"{u}neisen parameter.
(The relation between $\gamma_{\rm th}$ and the
thermodynamic Gr\"{u}neisen parameter $\gamma$ will be
discussed in Sec.~\ref{sec:other}.) We note
that if phonon dispersion curves at two different volumes
are related by a simple scaling factor, this must be the ratio
of $\bar{\omega}$ values at the two volumes. The scaling
factor used in Fig.~\ref{fig:phonons} was obtained in exactly this
way from our $\bar{\omega}$ results. The Gr\"{u}neisen parameter
$\gamma_{\rm ph}$ increases with increasing volume, in accord with
a widely used rule of thumb~\cite{anderson95}.
We find that $\gamma_{\rm ph}$
goes from 1.34 at $V = 6.7$~\AA$^3$ to 1.70 at $V = 8.3$~\AA$^3$,
but then decreases slightly to 1.62 at $V = 9.5$~\AA$^3$.
Fig.~\ref{fig:phonons}
also allows us to judge the effect of $T_{\rm el}$ on phonon
frequencies: for all volumes studied, the frequencies decrease
by {\em ca.}~4~\% as $T_{\rm el}$
goes from 2000 to 6000~K. However, we mention that
for the higher volumes, though not for the smaller ones,
$\bar{\omega}$ slightly increases again as $T_{\rm el}$
goes to still higher values.
To enable the $\bar{\omega}$ data to be used in thermodynamic
calculations, we parameterise the temperature
dependence of $\ln \bar{\omega}$ at each volume
as $a + b T^2 + c T^3 + e T^5$, and the volume dependence
of the four coefficients $a$, $b$, $c$ and $e$
as a third-degree polynomial in $V$.

We now return to the matter of quantum nuclear corrections. Since
the leading high-temperature correction to the free energy
is $\frac{1}{24} k_{\rm B} T ( \beta \hbar \omega )^2$ per
mode and there are three modes per atom, the quantum
correction to $F_{\rm harm}$ is $\frac{1}{8} k_{\rm B} T
( \beta \hbar \langle \omega^2 \rangle^{1/2} )^2$ per atom,
where $\langle \omega^2 \rangle$ denotes the average
of $\omega^2$ over wavevectors and branches. At the lowest
volume of interest, $V = 7$~\AA$^3$,
$\langle \omega^2 \rangle^{1/2} / 2 \pi$ is roughly 15~THz.
At the lowest temperature of interest, $T = 2000$~K, this gives
a quantum correction of 3~meV/atom, which is small compared
with our target precision.

\subsection{Harmonic phonon specific heat and thermal pressure}
\label{harmCp}

If the mean frequency $\bar{\omega}$ were independent of temperature,
the constant-volume specific heat $C_{\rm harm}$ due to harmonic
phonons would be exactly $3 k_{\rm B}$ per atom in the classical limit
employed here. We find that its temperature dependence
yields a slight increase of $C_{\rm harm}$ above this value, but
this is never greater than $0.25 k_{\rm B}$ under the conditions
of interest. The harmonic
phonon pressure $p_{\rm harm}$ as a function of atomic
volume at different temperatures is reported in
Fig.~\ref{fig:p_harmonic}. Comparison with Fig.~\ref{fig:p_electronic}
shows that $p_{\rm harm}$ is always much bigger (by a factor
of at least three) than the electronic thermal pressure under
the conditions of interest. At ICB conditions ($p = 330$~GPa,
$T \sim 5000 - 6000$~K), $p_{\rm harm}$ account for {\em ca.}~15~\%
of the total pressure.

\section{Anharmonic free energy}
\label{sec:anharmonic}

\subsection{Optimisation of reference system}
\label{sec:optimisation}
It was stressed in Sec.~\ref{sec:reference}
that optimisation of the reference
system greatly improves the efficiency of the anharmonic
calculations. We investigated the construction of the reference
system in detail at the atomic volume 8.67~\AA$^3$, with
the optimisations performed for a simulated system
of 16 atoms. The calculation of the anharmonic free energy
itself for a system as small as this would not be adequate,
but we expect this system size to suffice for the
optimisation of $U_{\rm ref}$. The initial sample of configurations
(see Sec.~\ref{sec:reference}) was taken from a
simulation of duration 100~ps
performed with the total energy $U_{\rm harm}$, with velocity
randomisation typically every 0.2~ps. Configurations were
taken every 1~ps, so that we obtain a sample of 100
configurations. In computing the energy difference
$U_{\rm vib} - U_{\rm ref}$ for these configurations,
the {\em ab initio} energy $U_{\rm vib}$ was always computed using
$5 \times 5 \times 3$ Monkhorst-Pack electronic
$k$-point sampling (38 $k$-points in the full Brillouin zone).
Once the preliminary optimisation
had been performed with configurations generated like
this, the resulting $U_{\rm ref}$ was used to produce a
new set of 100 configurations with an Andersen MD simulation of the
same duration as before, and the reference system was reoptimised.

This entire procedure was carried out at temperatures of 1000
and 4000~K. The values of the optimisation coefficients
(see Eqn~(\ref{eqn:optcoef})) were $c_1= 0.2$, $c_2 = 0.8$ at the
high temperature and $c_1 = 0.7$, $c_2 = 0.3$ at the low
temperature. (We do not require that $c_1 + c_2 = 1$,
though this happens to be the case here.) As expected, $U_{\rm ref}$
resembles $U_{\rm harm}$ quite closely at the
low temperature and $U_{\rm IP}$ quite closely at the
high temperature.

In view of the labour involved in the optimisation, we wanted to find out
whether the detailed choice of $c_1$ and $c_2$ makes a large
difference to the strength of the fluctuations of
$U_{\rm vib} - U_{\rm ref}$. To do this, we computed these
fluctuations at several temperatures, using the two
reference models just described, i.e. without optimising
the $c_i$ coefficients at each temperature. Our conclusion
is that the values $c_1 = 0.2$, $c_2 = 0.8$ can safely
be used at all the state points of interest, without
incurring large fluctuations, and we therefore
used this way of making the reference system in all
subsequent calculations.

\subsection{From harmonic {\em ab initio} to reference
to full {\em ab initio}}
\label{sec:harm2full}

The thermodynamic integration from {\em ab initio}
harmonic to reference was done with nine equally-spaced
$\lambda$-points using Simpson's rule, which gives
an integration precision well in excess of our target.
To investigate the influence of system size, integration
from $U_{\rm harm}$ to $U_{\rm ref}$ was performed
for systems of 12 different sizes, going from 16 to 1200 atoms.
The calculations were also repeated with
the force-constant matrix in $U_{\rm harm}$ generated
with cells containing from 16 to 150 atoms (see Sec.~\ref{sec:phonons}).
These tests showed that if the thermodynamic integration
is done with a system of 288 atoms and the force
constant used for $U_{\rm harm}$ is generated with
the 36-atom cell, 
then the resulting difference 
$F_{\rm ref} - F_{\rm harm}$ is converged to better than 3~meV/atom.

To compute the difference $F_{\rm vib} - F_{\rm ref}$, we
used the second-order expansion formula given
in Eqn~(\ref{eqn:secondorder}). Given
the small size of the fluctuations of $U_{\rm vib} -
U_{\rm ref}$, we expect this to be very accurate.
The calculations of $F_{\rm vib} - F_{\rm ref}$ were all done
with the 16-atom system. Tests with 36- and 64-atom systems show
that this free-energy difference is converged with respect
to size effects to within {\em ca.}~2~meV.

A summary of all our results for $F_{\rm ref} - F_{\rm harm}$
and $F_{\rm vib} - F_{\rm ref}$, and the resulting values
of $F_{\rm anharm} \equiv F_{\rm vib} - F_{\rm harm}$
are reported in Table~\ref{tab:anharm}.
The anharmonic free energy is always negative, so that
anharmonicity stabilises the solid. As expected, the anharmonic
free energy is small (less than or comparable with our target precision
of 10~meV/atom) at low temperatures, but increases rapidly
at high temperatures. The temperature at which it becomes appreciable
is higher for smaller atomic volumes.

In classical statistical mechanics, $F_{\rm anharm}$
is expected to go as $T^2$ at low temperatures, and in fact we find
that $F_{\rm anharm} = a(V) T^2$ gives a good
representation of our results for all the temperatures studied.
The volume dependence of $a(V)$ is adequately
represented by $a(V) = \alpha_1 + \alpha_2 V$, with
$\alpha_1 = 2.2 \times 10^{-9}$~eV~K$^{-2}$ and $\alpha_2 =
- 6.0 \times 10^{-10}$~eV~\AA$^{-3}$ per atom.

\subsection{Anharmonic specific heat and pressure}
\label{sec:anharmCp}

Within the parameterisation just described, the anharmonic
contribution to the constant-volume specific heat
$C_{\rm anharm}$ is proportional
to $T$ and varies linearly with $V$. As an indication of its
general size, we note that $C_{\rm anharm}$ increases from
0.09 to 0.18~$k_{\rm B}$ at 2000~K and from 0.28 to 0.53~$k_{\rm B}$
at 6000~K as $V$ goes from 7 to 10~\AA$^3$. The anharmonic
contribution to the pressure is independent of volume, and is
proportional to $T^2$. It increases from 0.4 to 3.5~GPa as $T$
goes from 2000 to 6000~K, so that even at high temperatures
it is barely significant.

\section{Thermodynamics of the solid}
\label{sec:thermo}

We now combine the parameterised forms for $F_{\rm perf}$,
$F_{\rm harm}$ and $F_{\rm anharm}$ presented in the previous
three Sections to obtain the total free energy of the
h.c.p. crystal, and hence, by taking appropriate derivatives,
a range of other thermodynamic functions, starting with those
measured in shock experiments.

\subsection{Thermodynamics on the Hugoniot}
\label{sec:hugoniot}
In a shock
experiment, conservation of mass, momentum and energy
require that the pressure $p_{\rm H}$, the molar internal
energy $E_{\rm H}$ and the molar volume $V_{\rm H}$ in the
compression wave are related by the
Rankine-Hugoniot formula~\cite{rankinehugoniot}:
\begin{equation}
\frac{1}{2} p_{\rm H} ( V_0 - V_{\rm H} ) =
E_{\rm H} - E_0 \; ,
\end{equation}
where $E_0$ and $V_0$ are the internal energy and volume in
the zero-pressure state before the arrival of the wave.
The quantities directly measured are the shock-wave and material
velocities, which allow the values of $p_{\rm H}$ and $V_{\rm H}$
to be deduced. From a series of experiments, $p_{\rm H}$ as
a function of $V_{\rm H}$ (the so-called Hugoniot) can
be derived. The measurement of temperature in shock experiments
is attempted but problematic~\cite{yoo93}.

The Hugoniot curve $p_{\rm H} ( V_{\rm H} )$ is straightforward
to compute from our results: for a given $V_{\rm H}$, one
seeks the temperature at which the Rankine-Hugoniot
relation is satisfied; from this, one
obtains $p_{\rm H}$ (and, if required, $E_{\rm H}$). In
experiments on Fe, $V_0$ and $E_0$ refer to the zero-pressure b.c.c.
crystal, and we obtain their values directly from GGA calculations,
using exactly the same PAW technique and GGA as in the rest
of the calculations. Since b.c.c. Fe is ferromagnetic,
spin polarisation must be included, and this is treated
by spin interpolation of the correlation energy due
to Vosko {\em et al.}, as described in
Refs.~\cite{alfe99b,kresse99}. The value of
$E_0$ includes the harmonic vibrational energy at 300~K, calculated
from {\em ab initio} phonon dispersion relations for
ferromagnetic b.c.c. Fe.

Our {\em ab initio} Hugoniot is compared with the measurements
of Brown and McQueen~\cite{brown86} in Fig.~\ref{fig:p_hugoniot}.
The agreement is
good, with discrepancies ranging from 10~GPa at $V = 7.8$~\AA$^3$
to 12~GPa at $V = 8.6$~\AA$^3$. These discrepancies are only slightly
greater than those found for the room-temperature static $p(V)$
curve (see Sec.~\ref{sec:perfect}), which can be
regarded as giving an indication of the
intrinsic accuracy of the GGA itself. Another way
of looking at the accuracy to be expected of the GGA is
to recalculate the Hugoniot using the experimental
value of the b.c.c. $V_0$ (11.8~\AA$^3$, compared with the
{\em ab initio} value of 11.55~\AA$^3$). The Hugoniot calculated
in this way is also plotted in Fig.~\ref{fig:p_hugoniot},
and we see that this
gives almost perfect agreement with the experimental data in
the pressure range $100 - 240$~GPa. We deduce from this
that the {\em ab initio} Hugoniot deviates from
the experimental data by an amount which should
be expected from the known inaccuracies of the GGA applied to Fe.
A similar comparison with the experimental Hugoniot
was given in the tight-binding total-energy work of Wasserman
{\em et al.}~\cite{wasserman96}, and their agreement was as good
as ours. We discuss the significance of this later.

Our Hugoniot temperature as a function of pressure is compared
with the experimental results of Brown and McQueen~\cite{brown86}
and of Yoo {\em et al.}~\cite{yoo93}
in Fig.~\ref{fig:t_hugoniot}. The {\em ab initio} temperatures
agree well with those of Brown and McQueen, but fall
substantially below those of Yoo {\em et al.}, and this
supports the suggestion of Ref.~\cite{wasserman96} that the Yoo {\em et al.}
measurements overestimate the Hugoniot temperature
by {\em ca.}~1000~K.

A further quantity that can be extracted from shock experiments
is the bulk sound velocity $v_{\rm B}$ as a function of
atomic volume on the Hugoniot, which is given by
$v_{\rm B} = ( K_S / \rho )^{1/2}$, with $K_S \equiv
- V ( \partial p / \partial V )_S$ the adiabatic
bulk modulus and $\rho$ the mass density. Since $K_S$ can be calculated
from our {\em ab initio} pressure and entropy as functions
of $V$ and $T$, our calculated $K_S$
can be directly compared with experimental
values (Fig.~\ref{fig:ks_hugoniot}). Here, there is a greater discrepancy
than one would wish, with the theoretical values falling
significantly above the $K_S$ values of both Refs~\cite{brown86}
and \cite{jeanloz79}, although we note that the two sets of
experimental results disagree by an amount comparable with the
discrepancy between theory and experiment.

For what it is worth, we show in Fig.~\ref{fig:alpha_hugoniot}
a comparison between our calculated thermal expansivity on the
Hugoniot with  values extracted from shock data by Jeanloz~\cite{jeanloz79}.
The latter are very scattered, but is clear that the theoretical
values have similar magnitude. However, our values vary little along
the Hugoniot, whereas the experimental values seem
to decrease rather rapidly with increasing pressure.

\subsection{Other thermodynamic quantities}
\label{sec:other}

We conclude our presentation of results by reporting our {\em ab
initio} predictions of quantities which conveniently
characterise h.c.p. Fe at high pressures and temperatures,
and allow some further comparisons with the predictions
of Refs~\cite{stixrude97} and \cite{wasserman96}. Our results
are presented as a function of pressure on isotherms
at $T = 2000$, 4000 and 6000~K. At each temperature, we give results only
for the pressure range where, according to our {\em ab initio}
melting curve, the h.c.p. phase is thermodynamically stable.

The total constant-volume specific heat per atom $C_v$
(Fig.~\ref{fig:cv_total}) emphasises again the importance
of electronic excitations. In a purely harmonic system,
$C_v$ would be equal to $3 k_{\rm B}$, and it is striking
that $C_v$ is considerably greater than that even at the modest
temperature of 2000~K, while at 6000~K it is
nearly doubled. The decrease of $C_v$ with increasing pressure
evident in Fig.~\ref{fig:cv_total} comes from the suppression
of electronic excitations by high compression, and to a smaller
extent from the suppression of anharmonicity.

The thermal expansivity $\alpha$ (Fig.~\ref{fig:alpha}) is one of the few
cases where we can compare with DAC
measurements~\cite{boehler90}. The latter
show that $\alpha$ decreases strongly with increasing pressure
and the {\em ab initio} results fully confirm this. Our results
also show that $\alpha$ increases significantly with temperature.
Both trends are also shown by the tight-binding calculations
of Ref.~\cite{wasserman96}, though the latter differ from
ours in showing considerably larger values of $\alpha$
at low pressures. We note that Ref.~\cite{wasserman96}
reported results for $\alpha$ at temperatures only
up to 2000~K, so a full comparison is not possible.

The product $\alpha K_T$ of expansivity and isothermal bulk
modulus is important because it is sometimes assumed to be
independent of pressure and temperature over a wide range
of conditions, and this constancy is used to extrapolate
experimental data. Our predicted isotherms for $\alpha K_T$
(Fig.~\ref{fig:alphakt}) indicate that its dependence on
$p$ is indeed weak, especially at low temperatures, but that its
dependence on $T$ certainly cannot be ignored, since it increases by at
least 30~\% as $T$ goes from 2000 to 6000~K at high pressures.
Wasserman {\em et al.}~\cite{wasserman96} come to
qualitatively similar conclusions, and they also find values
of {\em ca.}~10~MPa~K$^{-1}$ at $T \simeq 2000$~K. However, it
is disturbing to note that the general tendency for
$\alpha K_T$ to increase with pressure evident in our
results is exactly the opposite of what was found in Ref.~\cite{wasserman96}.
In particular, they found a marked increase of $\alpha K_T$ as
$p \rightarrow 0$, which does not occur in our results.

The thermodynamic Gr\"{u}neisen parameter $\gamma \equiv
V ( \partial p / \partial E )_V \equiv \alpha
K_T V / C_v$ plays an important role in high-pressure
physics, because it relates the thermal pressure (i.e. the
difference $p_{\rm th}$ between $p$ at given $V$ and $T$
and $p$ at the same $V$ but $T = 0$) and the thermal energy
(difference $E_{\rm th}$ between $E$ at given $V$ and $T$ and $E$
at the same $V$ but $T = 0$). Assumptions about the value of $\gamma$ are
frequently used in reducing shock data from Hugoniot
to isotherm. If one assumes that $\gamma$ depends only on $V$, then
the thermal pressure and energy are related by:
\begin{equation}
p_{\rm th} V = \gamma E_{\rm th} \; ,
\end{equation}
a relation known as the Mie-Gr\"{u}neisen equation of state. At
low temperatures, where only harmonic phonons contribute
to $E_{\rm th}$ and $p_{\rm th}$, $\gamma$ should indeed be
temperature independent above the Debye temperature,
because $E_{\rm th} = 3 k_{\rm B} T$ per atom, and
$p_{\rm th} V = - 3 k_{\rm B} T d \ln \bar{\omega} /
d \ln V = 3 k_{\rm B} T \gamma_{\rm th}$, so that
$\gamma = \gamma_{\rm ph}$, which depends only on $V$.
But in high-temperature Fe, the temperature independence
of $\gamma$ will clearly fail, because of electronic
excitations (and anharmonicity).

Our results for $\gamma$ (Fig.~\ref{fig:gamma}) indicate that
it varies rather little with either pressure or temperature
in the region of interest. At temperatures below {\em ca.}~4000~K,
it decreases with increasing pressure, as expected from the
behaviour of the phonon Gr\"{u}neisen parameter $\gamma_{\rm ph}$ (see
Sec.~\ref{sec:dispersion}). This is also
expected from the often-used empirical
rule of thumb~\cite{anderson95} $\gamma \simeq ( V / V_0 )^q$, where
$V_0$ is a reference volume and $q$ is a constant exponent
usually taken to be roughly unity. Since $V$ decreases
by a factor of about 0.82 as $p$ goes from 100 to 300~GPa, this empirical
relation would make $\gamma$ decrease by the same factor
over this range, which is roughly what we see. However, the pressure
dependence of $\gamma$ is very much weakened as $T$ increases,
until at 6000~K $\gamma$ is almost constant.

Our results agree quite well with those of Wasserman
{\em et al.}~\cite{wasserman96} in giving a value $\gamma \simeq 1.5$
at high pressures, although once again their calculations
are limited to the low-temperature region $T \le 3000$~K. But
at low pressures there is a serious disagreement, since they find
a strong increase of $\gamma$ to values of well over 2.0 as
$p \rightarrow 0$, whereas our values never exceed 1.6.

\section{Discussion and conclusions}
\label{sec:discon}

Our primary interest in this work is in the properties of
h.c.p. iron at high pressures and temperatures, but in order
to investigate them using {\em ab initio} methods we have needed
to make technical developments, which have a wider significance.
The major technical achievement is that we have been able
to calculate the {\em ab initio} free energy and other thermodynamic
properties with
completely controlled statistical-mechanical errors, i.e. errors
that can be reduced to any required extent. Anharmonicity
and thermal electronic excitations are fully included.
The attainment of high precision for the electronic
and harmonic parts of the free energy has required no
particular technical innovations, though careful attention
to sources of error is essential. The main innovation is
in the development of well optimised reference systems
for use with thermodynamic integration in the calculation
of the anharmonic part, without which adequate precision would
be impossible. With the methods we have developed, it becomes
unnecessary to approximate the electronic structure with semi-empirical
representations, or to resort to the statistical-mechanical
approximations that have been used in the past.

We have assessed in detail the precision achieved in the various
parts of the free energy. There are two kinds of errors: those
incurred in the calculation of the free energies themselves,
and those produced by fitting the results to polynomials. We have seen
that the errors in calculating the perfect-lattice free energy
$F_{\rm perf}$ are completely negligible, though there may be
small fitting errors of perhaps 1~meV/atom. In the harmonic part
$F_{\rm harm}$, the calculational errors are {\em ca.}~3~meV/atom, most
of which comes from spatial truncation of the force-constant
matrix; the fitting error for $F_{\rm harm}$ are of about the same
size. The most serious errors are in the anharmonic part
$F_{\rm anharm}$, and these are {\em ca.}~5~meV/atom in the calculation
and {\em ca.}~4~meV/atom in the fitting. The overall technical errors
therefore amount to {\em ca.}~15~meV/atom, which is slightly
larger than our target of 10~meV/atom.

We stress that the precision just quoted does not take into
account errors incurred in the particular
implementation of DFT (PAW in the present work), for example
the error associated with the chosen split between valence
and core states. Such errors can in principle be systematically
reduced, but we have not attempted this here. Nor does it
account for the inaccuracy of the chosen $E_{\rm xc}$, or for the
neglect of  the temperature dependence of $E_{\rm xc}$. We shall
attempt to assess errors of this type in our separate paper on
the melting properties of Fe.

The most direct way to test the reliability of our methods is comparison
with shock data for $p(V)$ on the Hugoniot~\cite{brown86},
so it is gratifying
to find close agreement over the pressure range of interest. The
closeness of this agreement is inherently limited by the known
inaccuracies of the GGA employed, and we have shown that the discrepancies
are of the expected size. An important prediction of the
calculations is the temperature $T(p)$ on the Hugoniot,
since temperature is notoriously difficult to obtain in
shock experiments. Our results support the reliability
of the shock temperatures estimated by 
Brown and McQueen~\cite{brown86,not_measured}, and,
in agreement with Wasserman {\em et al.}~\cite{wasserman96},
we find that the temperatures of Yoo {\em et al.}~\cite{yoo93}
are too high by as much as 1000~K. This incidentally
lends support to the reliability of the Brown and McQueen
estimate of {\em ca.}~5500~K for the melting temperature
of Fe at 243~GPa. The situation is not so satisfactory
for the adiabatic bulk modulus $K_S$ on the Hugoniot,
since our {\em ab initio} values seem to be {\em ca.}~8~\% above
the shock values. But it should be remembered that even
at ambient conditions {\em ab initio} and experimental bulk
moduli frequently differ by this amount. The difficulties
may be partly on the experimental side, since even for b.c.c.
Fe at ambient conditions, experimental $K_S$ values span a
range of 8~\%.

Our calculations fully confirm the strong influence of electronic
thermal excitations~\cite{boness90,wasserman96}.
At the temperatures $T \sim 6000$~K of interest
for the Earth's core, their contribution to the specific
heat is almost as large as that due to lattice vibrations, in
line with previous estimates. They also have a significant
effect on the Gr\"{u}neisen parameter $\gamma$, which plays
a key role in the thermodynamics of the core, and is poorly
constrained by experiment. Our finding that $\gamma$ decreases
with increasing pressure for $T < 4000$~K accords with an often-used
rule of thumb~\cite{anderson95}, but electronic excitations completely
change this behaviour at core temperatures $T \sim 6000$~K, where
$\gamma$ has almost constant values of {\em ca.}~1.45,
in accord with experimental estimates in the range
1.1 to 1.6~\cite{brown86,stacey95}. Comparison with the earlier tight-binding
calculations of Wasserman {\em et al.}~\cite{wasserman96}
both for $\gamma$ and for the quantity $\alpha K_T$
is rather disquieting. Although a full
comparison is hindered by the fact that they report results
only for the low-temperature region $T \le 3000$~K, we find
two kinds of disagreement at low pressure. First, they find an
increase of $\alpha K_T$ as $p \rightarrow 0$, whreas at
low temperatures we find the opposite. Even more seriously,
their strong increase of $\gamma$ as $p \rightarrow 0$
is completely absent in our results. Our calculations
are more rigorous than theirs, since we completely avoid
their statistical-mechanical approximations, as well as being
fully self-consistent on the electronic-structure side. The
suggestion must be that their approximations lead to significantly
erroneous behaviour at low pressures. In pursuing this further,
it would be very helpful to know what their
methods predict in the high-temperature region relevant
to the Earth's core.

The present work forms part of a larger project on both pure Fe
and its alloys with S, O, Si and H in the solid and liquid
states. In a separate paper, we shall demonstrate
that the thermodynamic integration technique employed here
can also be used to obtain the fully {\em ab initio}
free energy and other thermodynamic functions of liquid Fe over
a wide range of states, with a precision equal to what has
been achieved here for the solid. From the free energies of
solid and liquid, we are then able to determine the {\em ab initio}
melting curve and the entropy and volume of fusion as functions
of pressure.

In summary, we have presented extensive {\em ab initio}
calculations of the free energy and a range of
other thermodynamic properties of iron at high
pressures and temperatures, in which all statistical-mechanical
errors are fully under control, and a high (and quantified)
precision has been achieved. We find close agreement
with the most reliable shock data.
{\em Ab initio} values are provided for important, but
experimentally poorly determined quantities, such as the
Gr\"{u}neisen parameter. The free energy results provide
part of the basis for the {\em ab initio} determination of
the high-pressure melting properties of iron, to be
reported elsewhere.

\section*{Acknowledgments}
The work of DA is supported by NERC grant GST/02/1454 to G. D. Price
and M. J. Gillan. We thank NERC and EPSRC for allocations of
time on the Cray T3E machines at Edinburgh Parallel Computer Centre
and Manchester CSAR service, these allocations being provided
through the Minerals Physics Consortium (GST/02/1002) and the
UK Car-Parrinello Consortium (GR/M01753). We gratefully acknowledge
discussions with Prof. J.-P. Poirier and Dr. L. Vo\v{c}adlo.

\clearpage

\begin{table}
\begin{tabular}{lcccc}
$V$(\AA$^3$) & $T$(K) & $ F_{\rm ref} - F_{\rm harm} $ (eV) &
$ F_{\rm vib} - F_{\rm ref} $ (eV) &
$ F_{\rm vib} - F_{\rm harm} $ (eV) \\
\hline
9.17       & 1000 &     4.448     &       -4.455     & -0.007 \\
           & 1500 &     4.437     &       -4.449     & -0.012 \\
           & 2500 &     4.416     &       -4.440     & -0.024 \\
           & 3000 &     4.405     &       -4.441     & -0.036 \\
           & 3500 &     4.403     &       -4.444     & -0.041 \\
           & 4000 &     4.402     &       -4.453     & -0.051 \\
           & 4500 &     4.385     &       -4.456     & -0.071 \\
& & & & \\
8.67       & 1000 &     4.961     &       -4.968     & -0.007 \\
           & 2000 &     4.933     &       -4.950     & -0.017 \\
           & 3000 &     4.913     &       -4.941     & -0.028 \\
           & 3500 &     4.906     &       -4.941     & -0.035 \\
           & 4000 &     4.902     &       -4.939     & -0.042 \\
           & 4500 &     4.893     &       -4.951     & -0.058 \\
           & 5000 &     4.888     &       -4.954     & -0.067 \\
& & & & \\
8.06       & 2250 &     5.683     &       -5.704     & -0.021 \\
           & 3400 &     5.655     &       -5.687     & -0.032 \\
           & 4500 &     5.634     &       -5.691     & -0.057 \\
           & 5000 &     5.626     &       -5.687     & -0.061 \\
& & & & \\
7.5        & 2500 &     6.530     &       -6.554     & -0.024 \\
           & 4000 &     6.490     &       -6.531     & -0.041 \\
           & 5000 &     6.468     &       -6.525     & -0.057 \\
           & 6000 &     6.448     &       -6.533     & -0.085 \\
& & & & \\
6.97       & 3000 &     7.531     &       -7.549     & -0.018 \\
           & 4500 &     7.483     &       -7.522     & -0.039 \\
           & 5500 &     7.460     &       -7.523     & -0.063 \\
           & 6000 &     7.449     &       -7.514     & -0.065 \\
           & 6500 &     7.438     &       -7.519     & -0.081 \\
           & 7000 &     7.428     &       -7.523     & -0.095 \\
& & & & \\
\hline
\end{tabular}
\caption{Anharmonic contribution $F_{\rm anharm} \equiv
F_{\rm vib} - F_{\rm harm}$ to the {\em ab initio} free energy
of h.c.p. Fe at a set of atomic volumes $V$ and temperatures $T$,
calculated as a sum of the free energy difference $F_{\rm ref} -
F_{\rm harm}$ between reference and {\em ab initio} systems
and the difference $F_{\rm vib} - F_{\rm ref}$ between full
{\em ab initio} and reference systems. All quantities are per atom.}
\label{tab:anharm}
\end{table}

\clearpage

\begin{figure}
\caption{Electronic specific heat per atom $C_{\rm perf}$ of the rigid
perfect lattice of h.c.p. Fe (units of Boltzmann's constant $k_{\rm
B}$) as a function of temperature for atomic volumes: 7.0~\AA$^3$
(\protect\rule[1mm]{5mm}{0.1mm}), 8.0~\AA$^3$
(\protect\rule[1mm]{1.5mm}{.1mm} \protect\rule[1mm]{1.5mm}{.1mm}
\protect\rule[1mm]{1.5mm}{.1mm}), 9.0~\AA$^3$
(\protect\rule[1mm]{.5mm}{.1mm} \protect\rule[1mm]{.5mm}{.1mm}
\protect\rule[1mm]{.5mm}{.1mm}) and 10.0~\AA$^3$
(\protect\rule[1mm]{.1mm}{.2mm} \protect\rule[1mm]{.1mm}{.2mm}
\protect\rule[1mm]{.1mm}{.2mm}).}
\label{fig:cv_electronic}
\end{figure}

\begin{figure}
\caption{Electronic density of states of h.c.p. Fe calculated at
atomic volumes of 7.0~\AA$^3$ (\protect\rule[1mm]{5mm}{0.1mm}) and
10.0~\AA$^3$(\protect\rule[1mm]{.1mm}{.2mm}
\protect\rule[1mm]{.1mm}{.2mm} \protect\rule[1mm]{.1mm}{.2mm}).
Energy is referred to the Fermi energy $E_{\rm F}$.}
\label{fig:dos}
\end{figure}

\begin{figure}
\caption{Electronic thermal pressure $\Delta p_{\rm perf}$ of the
rigid perfect lattice of h.c.p. Fe as a function
of atomic volume $V$ for $T = 2000$ (\protect\rule[1mm]{5mm}{0.1mm}),
 4000 (\protect\rule[1mm]{1.5mm}{.1mm} \protect\rule[1mm]{1.5mm}{.1mm}
\protect\rule[1mm]{1.5mm}{.1mm}) and
6000~K (\protect\rule[1mm]{.5mm}{.1mm} \protect\rule[1mm]{.5mm}{.1mm}
\protect\rule[1mm]{.5mm}{.1mm}).}
\label{fig:p_electronic}
\end{figure}

\begin{figure}
\caption{Phonon dispersion relations of h.c.p. Fe calculated
at atomic volumes $V = 8.67$ (left panel) and 6.97~\AA$^3$
(right panel). Frequencies calculated directly from DFT
at the two volumes are shown as solid curves. In left panel,
dashed curves give frequencies from empirical inverse-power
model (see text). In right panel, dashed curves show
DFT frequencies for $V = 8.67$~\AA$^3$ graphed in left
panel but scaled by the factor 1.409.}
\label{fig:phonons}
\end{figure}

\begin{figure}
\caption{Geometric-mean phonon frequency $\bar{\omega}$
of h.c.p. Fe as a function of atomic volume $V$ for
$T = 2000$ (\protect\rule[1mm]{5mm}{0.1mm}),
 4000 (\protect\rule[1mm]{1.5mm}{.1mm} \protect\rule[1mm]{1.5mm}{.1mm}
\protect\rule[1mm]{1.5mm}{.1mm}) and
6000~K (\protect\rule[1mm]{.5mm}{.1mm} \protect\rule[1mm]{.5mm}{.1mm}
\protect\rule[1mm]{.5mm}{.1mm}). The natural
logarithm of the two quantities is plotted, with $\bar{\omega}$
in units of rad~s$^{-1}$ and $V$ in units of \AA$^3$.}
\label{fig:omega}
\end{figure}

\begin{figure}
\caption{The harmonic thermal pressure $p_{\rm harm}$ as
a function of atomic volume $V$ for 
$T = 2000$ (\protect\rule[1mm]{5mm}{0.1mm}),
 4000 (\protect\rule[1mm]{1.5mm}{.1mm} \protect\rule[1mm]{1.5mm}{.1mm}
\protect\rule[1mm]{1.5mm}{.1mm}) and
6000~K (\protect\rule[1mm]{.5mm}{.1mm} \protect\rule[1mm]{.5mm}{.1mm}
\protect\rule[1mm]{.5mm}{.1mm}).}
\label{fig:p_harmonic}
\end{figure}

\begin{figure}
\caption{Experimental and {\em ab initio} Hugoniot pressure $p$
as a function of atomic volume $V$. Symbols show the measurements
of Brown and McQueen~\protect\cite{brown86}. Solid curve is {\em ab initio}
pressure obtained when calculated equilibrium volume of
b.c.c. Fe is used in the Hugoniot-Rankine equation; dotted curve
is the same, but with experimental equilibrium volume
of b.c.c. Fe.}
\label{fig:p_hugoniot}
\end{figure}

\begin{figure}
\caption{Experimental and {\em ab initio} temperature as a function of
pressure on the Hugoniot. Black circles with error bars and white
diamonds are measurements due to Yoo {\em et al.}~\protect\cite{yoo93} and
Brown and McQueen~\protect\cite{brown86} respectively. Solid and
dashed curves are {\em ab initio} results obtained using theoretical
and experimental b.c.c. volumes.}
\label{fig:t_hugoniot}
\end{figure}

\begin{figure}
\caption{Experimental and {\em ab initio} adiabatic
bulk modulus $K_S$ on the Hugoniot. Diamonds and pluses are
measurements due to Jeanloz~\protect\cite{jeanloz79} and 
Brown and McQueen~\protect\cite{brown86} respectively. Solid
and dashed curves are {\em ab initio} results obtained using
theoretical and experimental b.c.c. volumes.}
\label{fig:ks_hugoniot}
\end{figure}

\begin{figure}
\caption{Experimental and {\em ab initio} thermal expansivity
on the Hugoniot. Diamonds are measurements due to 
Jeanloz~\protect\cite{jeanloz79}. Solid
and dashed curves are {\em ab initio} results obtained
using theoretical and experimental b.c.c. volumes.}
\label{fig:alpha_hugoniot}
\end{figure}

\begin{figure}
\caption{Total constant-volume specific heat per atom
$C_v$ (units of $k_{\rm B}$) of h.c.p Fe as a function of
pressure from present {\em ab initio}
calculations at 
$T = 2000$ (\protect\rule[1mm]{5mm}{0.1mm}),
4000 (\protect\rule[1mm]{1.5mm}{.1mm} \protect\rule[1mm]{1.5mm}{.1mm}
\protect\rule[1mm]{1.5mm}{.1mm}) and
6000~K (\protect\rule[1mm]{.5mm}{.1mm} \protect\rule[1mm]{.5mm}{.1mm}
\protect\rule[1mm]{.5mm}{.1mm}).}
\label{fig:cv_total}
\end{figure}

\begin{figure}
\caption{{\em Ab initio} thermal expansivity $\alpha$ as a
function of pressure on isotherms 
$T = 2000$ (\protect\rule[1mm]{5mm}{0.1mm}),
 4000 (\protect\rule[1mm]{1.5mm}{.1mm} \protect\rule[1mm]{1.5mm}{.1mm}
\protect\rule[1mm]{1.5mm}{.1mm}) and
6000~K (\protect\rule[1mm]{.5mm}{.1mm} \protect\rule[1mm]{.5mm}{.1mm}
\protect\rule[1mm]{.5mm}{.1mm}).
Black circle with error bar is
experimental value of Duffy and Ahrens~\protect\cite{duffy93} 
at $T = 5200 \pm 500$~K. Diamonds
are DAC values due to Boehler~\protect\cite{boehler90} for
temperatures between 1500 and 2000~K.}
\label{fig:alpha}
\end{figure}

\begin{figure}
\caption{{\em Ab initio} values of product of thermal expansion
coefficient $\alpha$ and isothermal bulk modulus $K_T$ as
a function of pressure $P$ at $T = 2000$ 
(\protect\rule[1mm]{5mm}{0.1mm}),
4000 (\protect\rule[1mm]{1.5mm}{0.1mm} \protect\rule[1mm]{1.5mm}{.1mm}
\protect\rule[1mm]{1.5mm}{.1mm}) and
6000~K (\protect\rule[1mm]{.5mm}{.1mm} \protect\rule[1mm]{.5mm}{.1mm}
\protect\rule[1mm]{.5mm}{.1mm}).}
\label{fig:alphakt}
\end{figure}

\begin{figure}
\caption{{\em Ab initio} Gr\"{u}neisen parameter $\gamma$
as a function of pressure at 
$T = 2000$ (\protect\rule[1mm]{5mm}{0.1mm}),
 4000 (\protect\rule[1mm]{1.5mm}{.1mm} \protect\rule[1mm]{1.5mm}{.1mm}
\protect\rule[1mm]{1.5mm}{.1mm}) and
6000~K (\protect\rule[1mm]{.5mm}{.1mm} \protect\rule[1mm]{.5mm}{.1mm}
\protect\rule[1mm]{.5mm}{.1mm}).}
\label{fig:gamma}
\end{figure}

\clearpage
\centerline{FIGURE 1}
\centerline{\psfig{figure=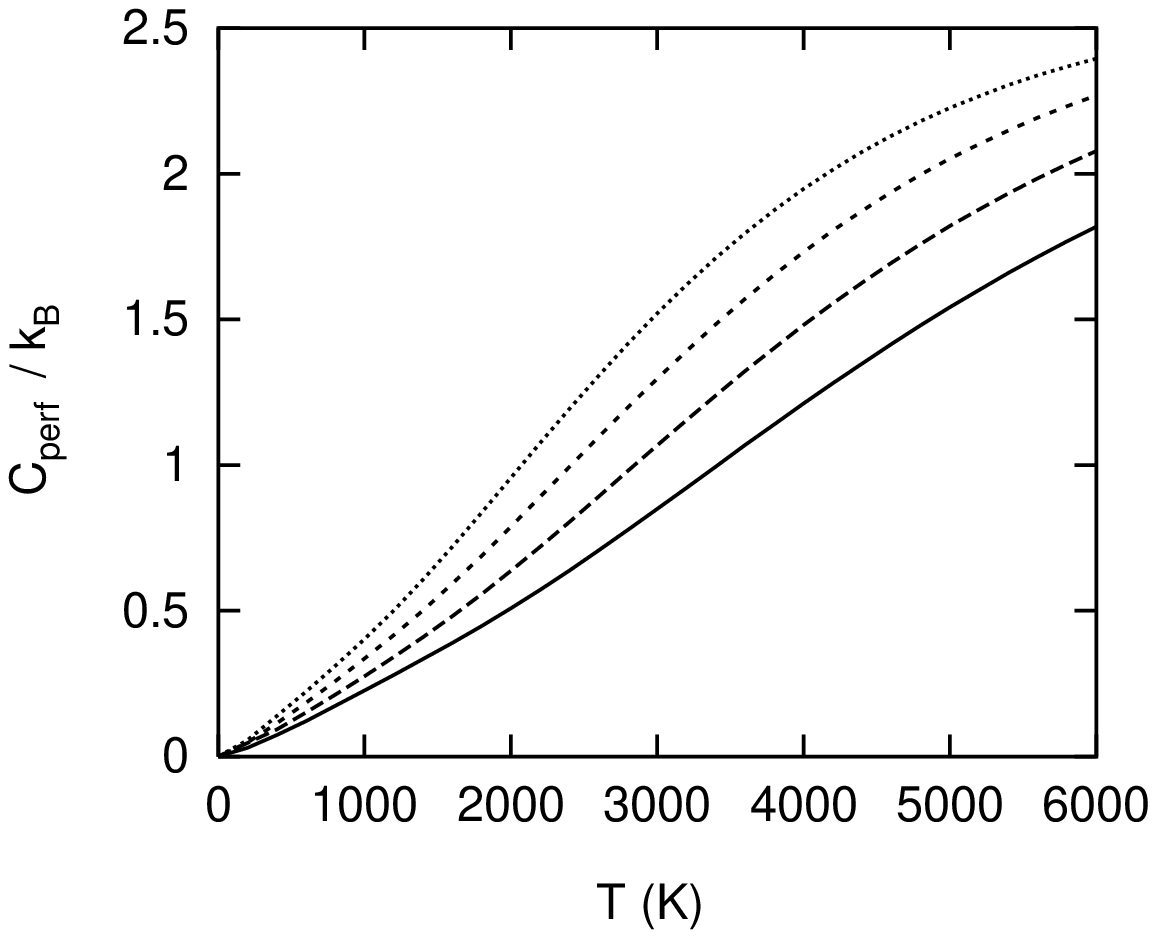,height=3in}}
\clearpage
\centerline{FIGURE 2}
\centerline{\psfig{figure=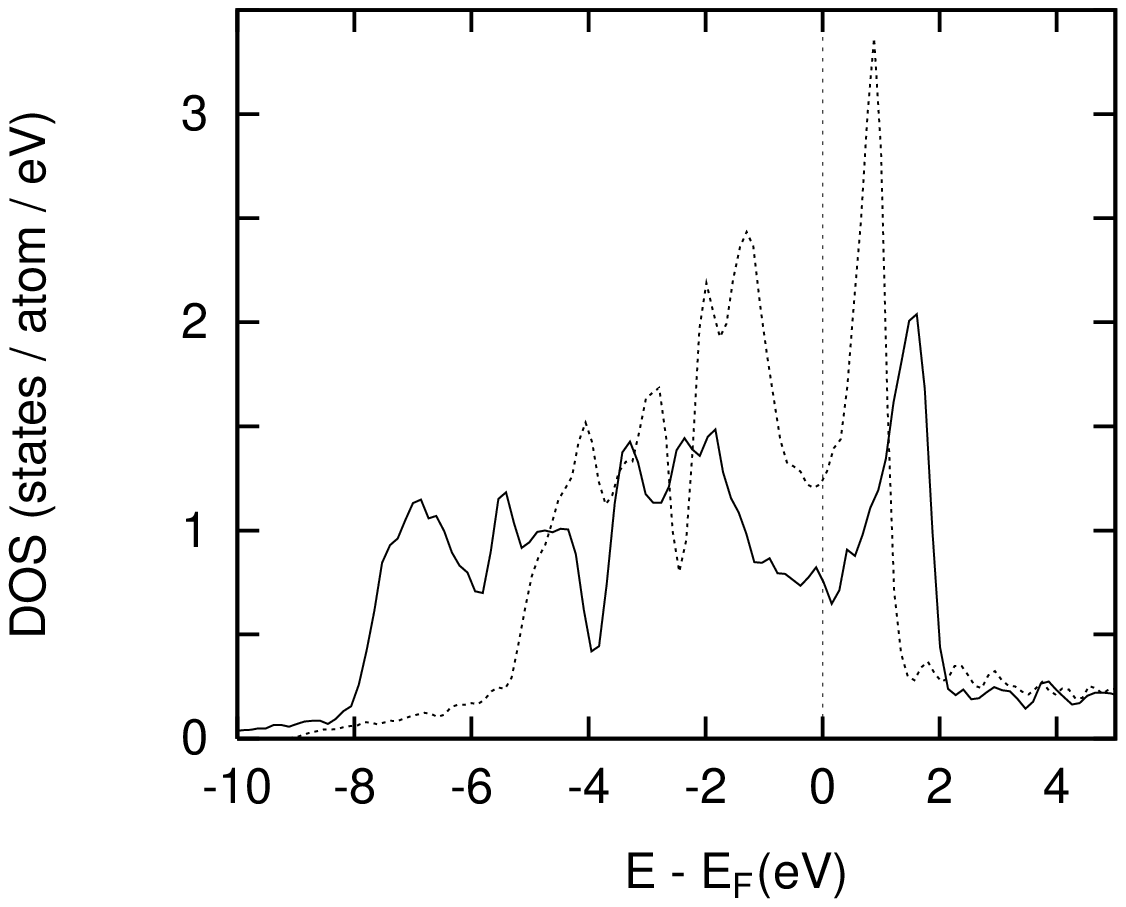,height=3in}}
\clearpage
\centerline{FIGURE 3}
\centerline{\psfig{figure=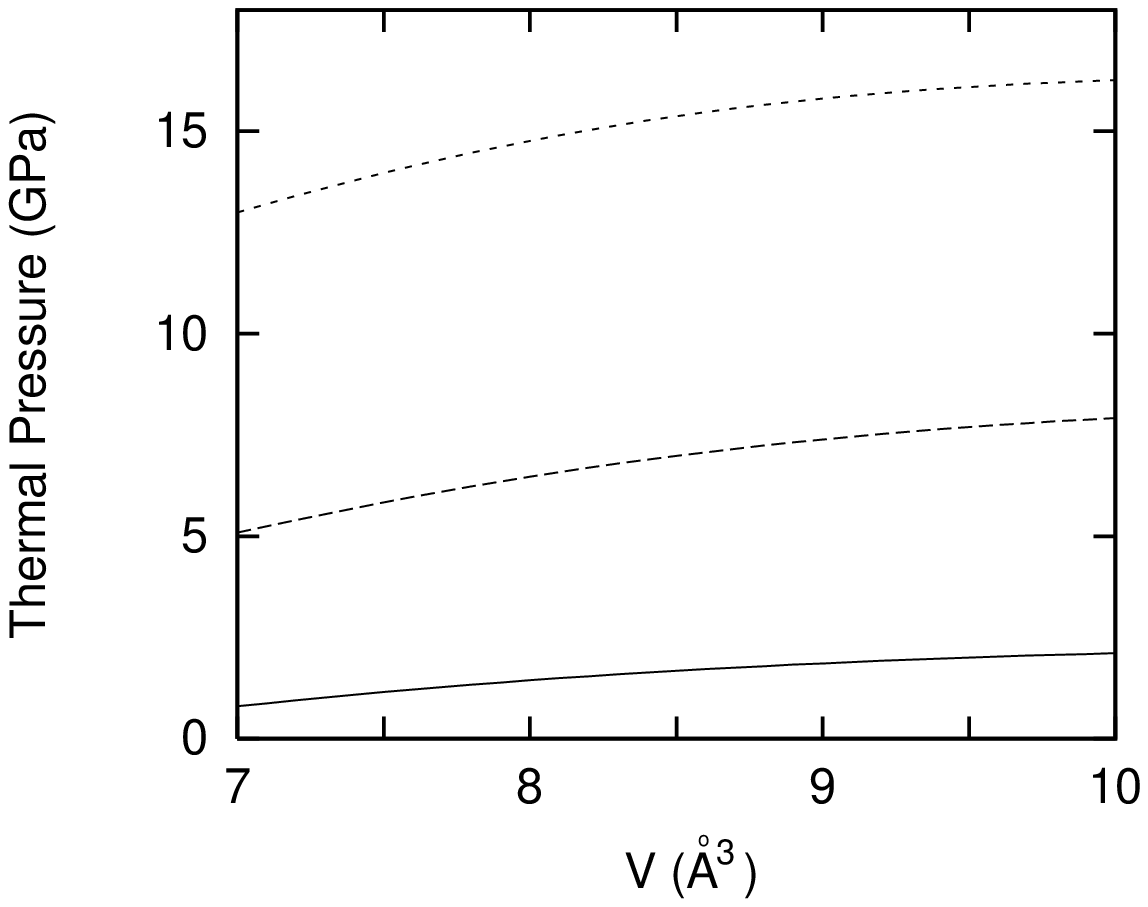,height=3in}}
\clearpage
\centerline{FIGURE 4}
\centerline{\psfig{figure=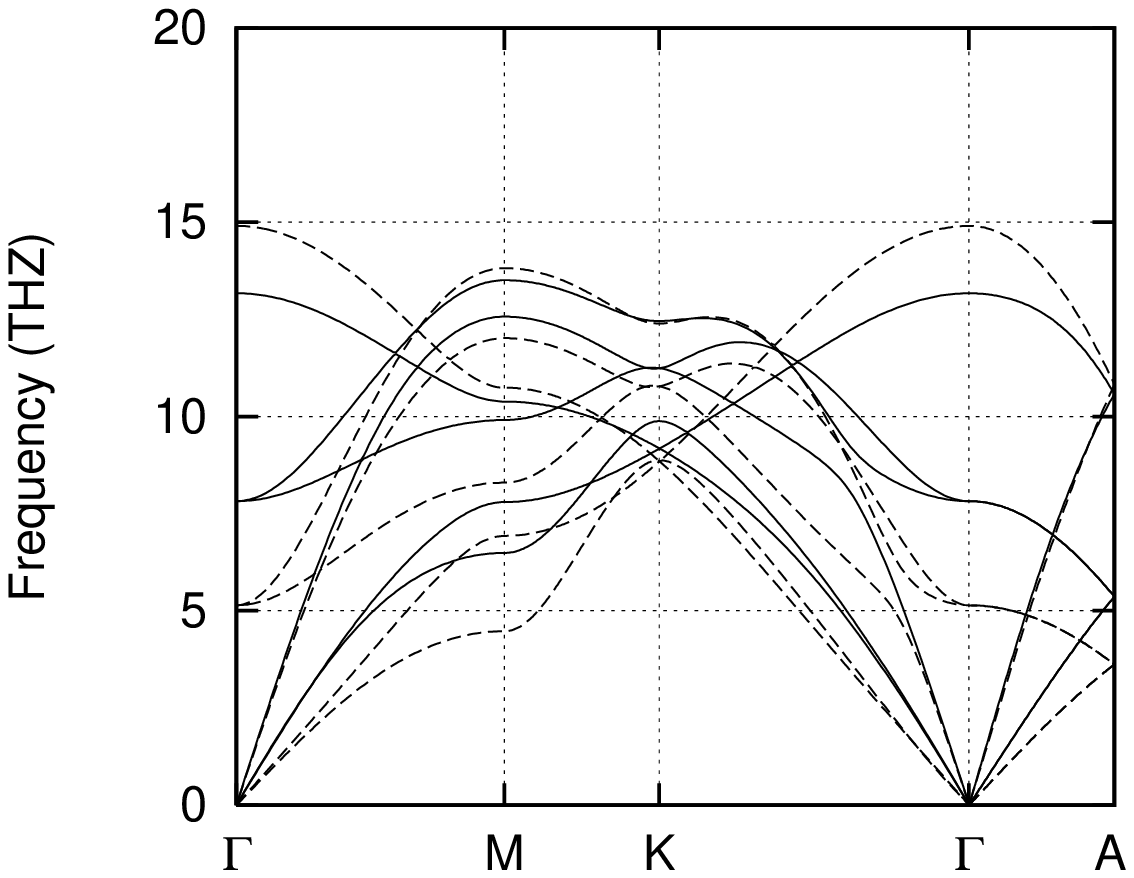,height=3in}
\psfig{figure=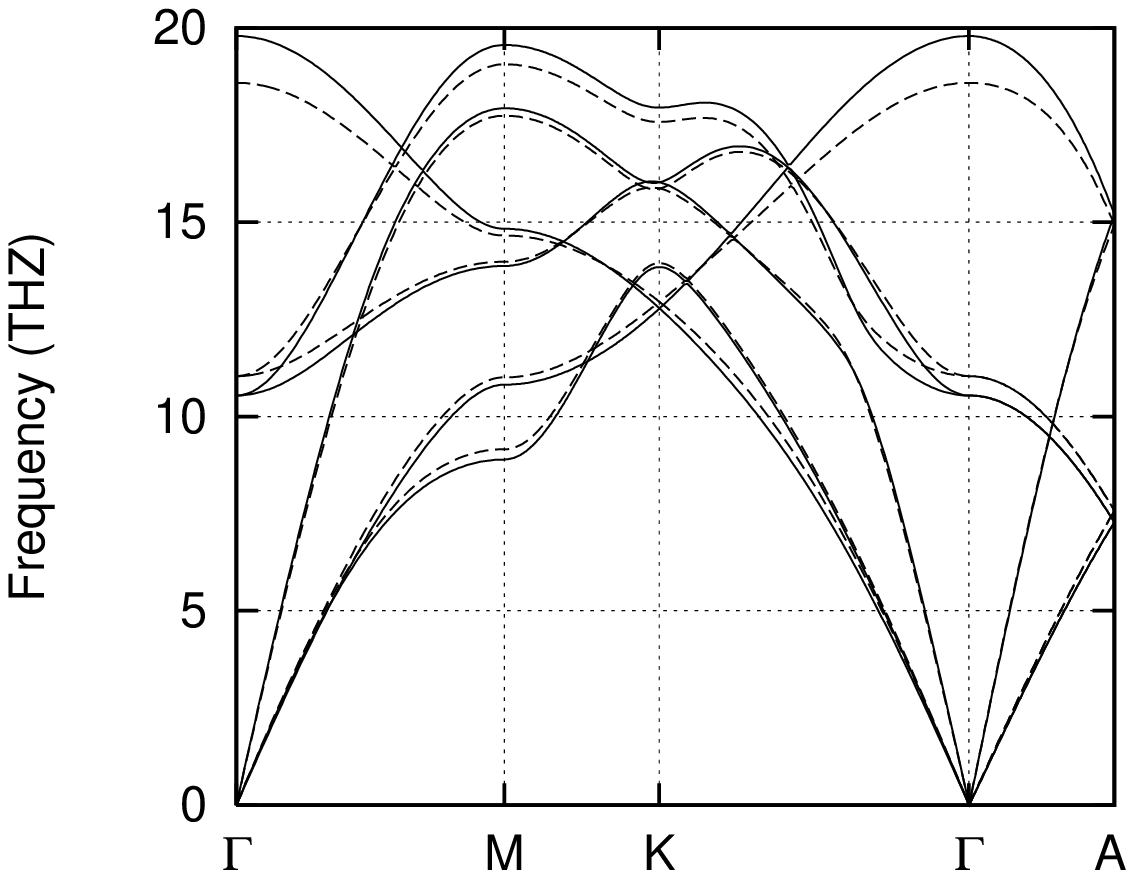,height=3in}}
\clearpage
\centerline{FIGURE 5}
\centerline{\psfig{figure=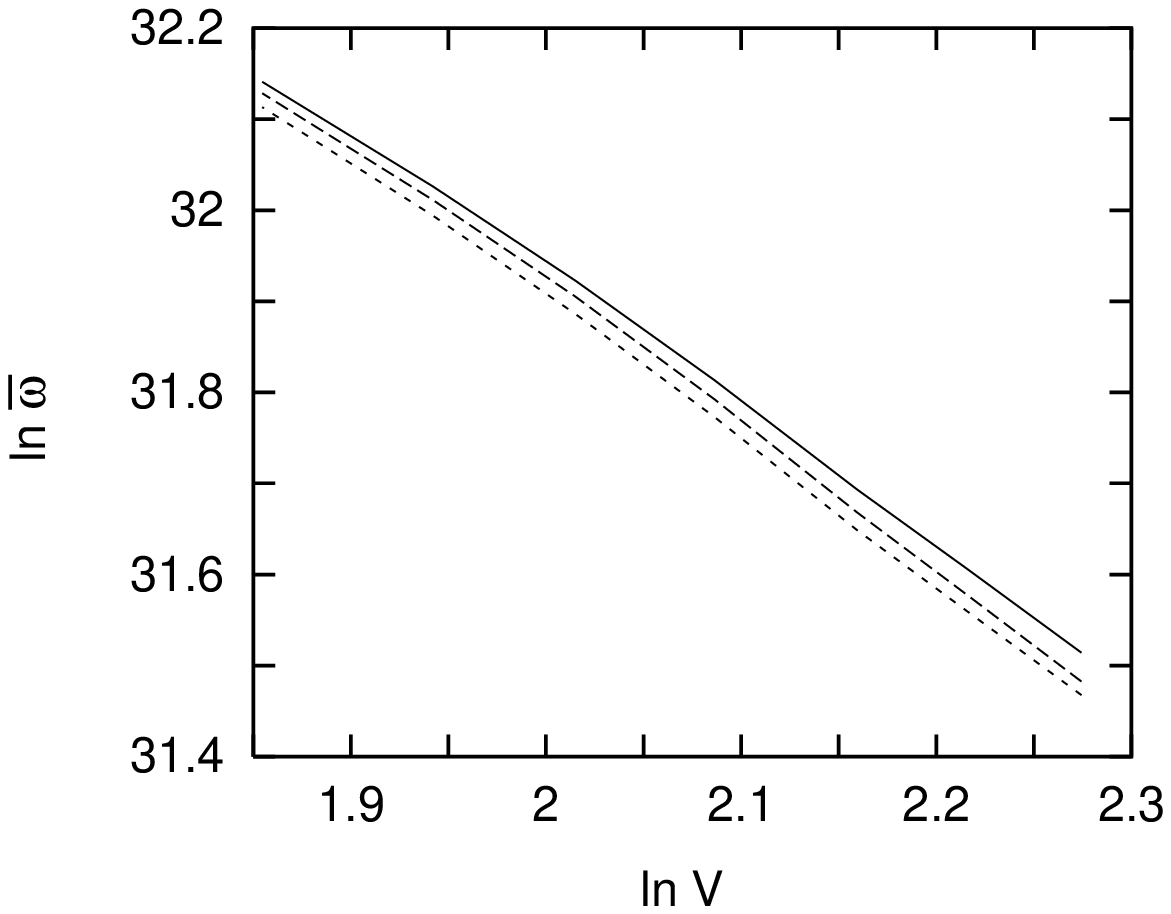,height=3in}}
\clearpage
\centerline{FIGURE 6}
\centerline{\psfig{figure=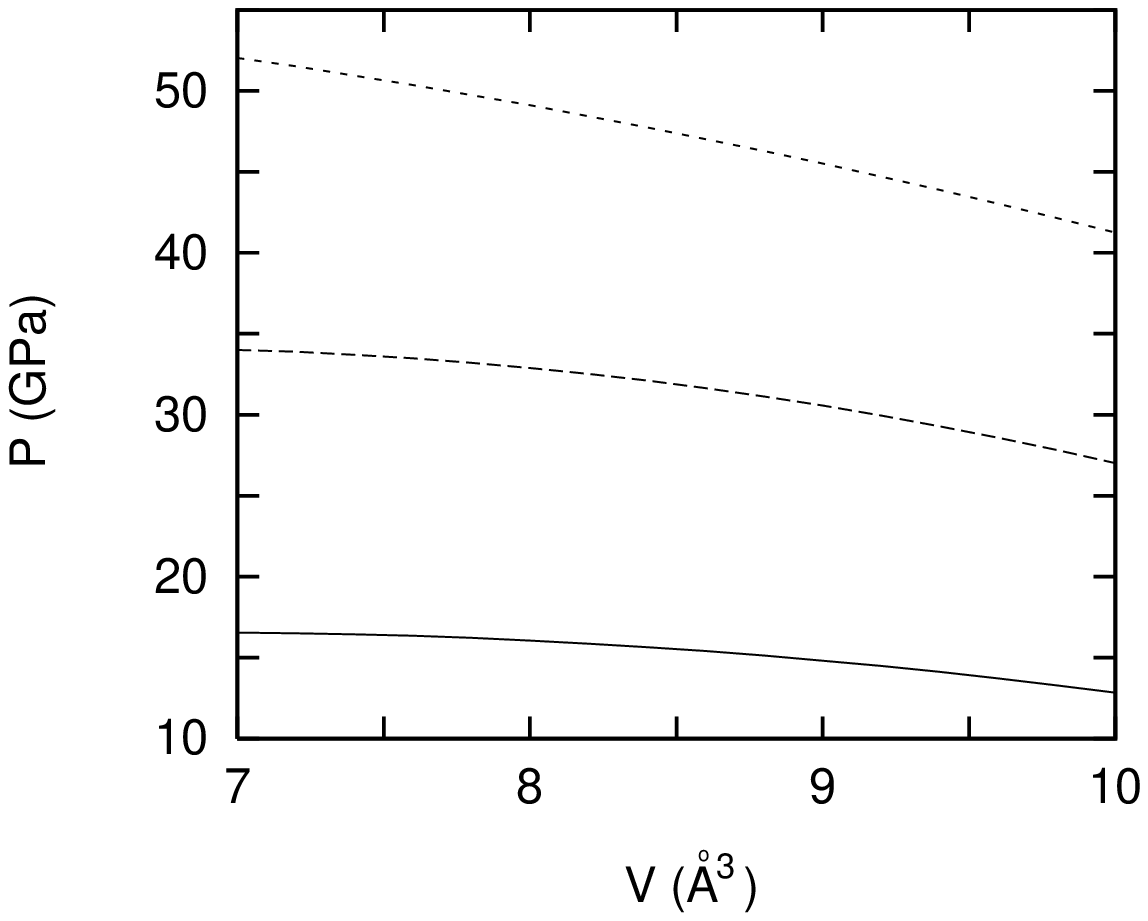,height=3in}}
\clearpage
\centerline{FIGURE 7}
\centerline{\psfig{figure=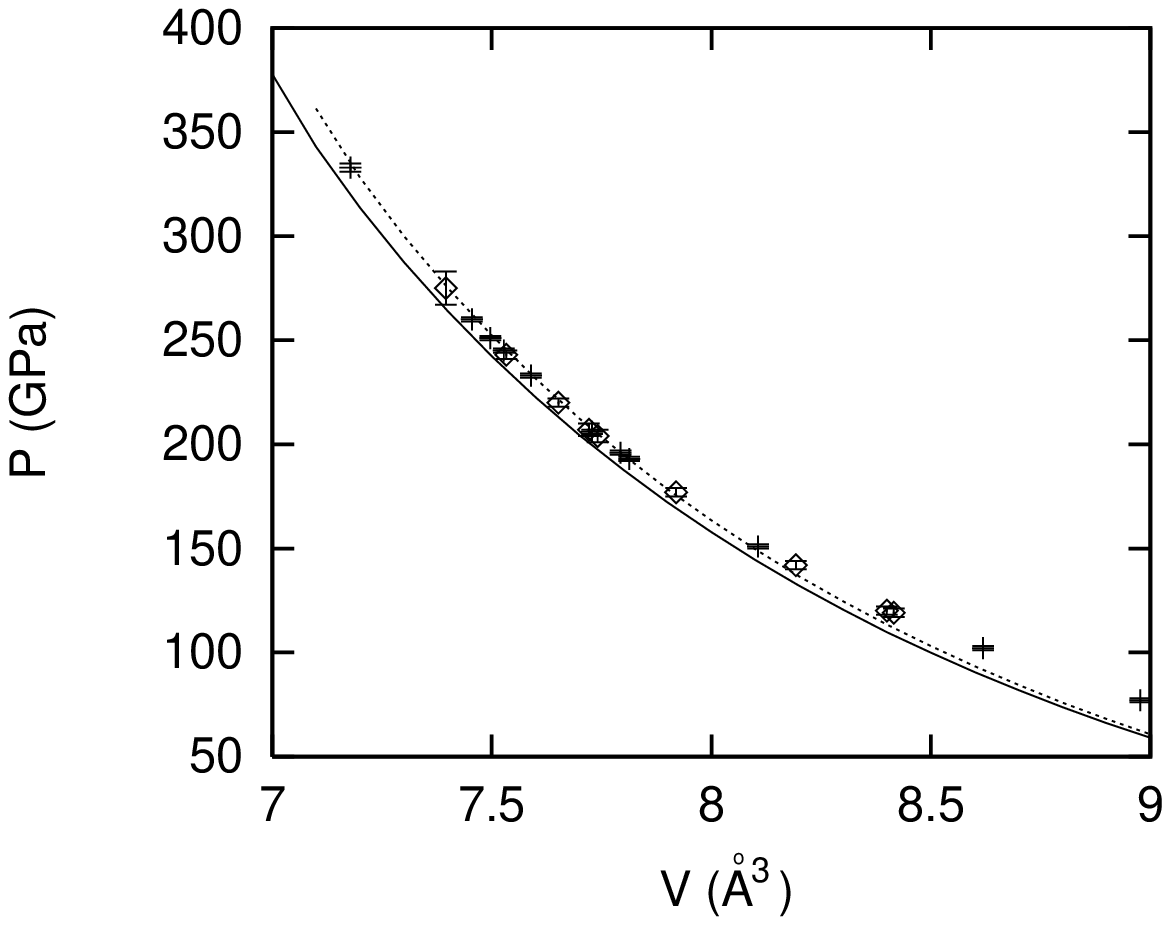,height=3in}}
\clearpage
\centerline{FIGURE 8}
\centerline{\psfig{figure=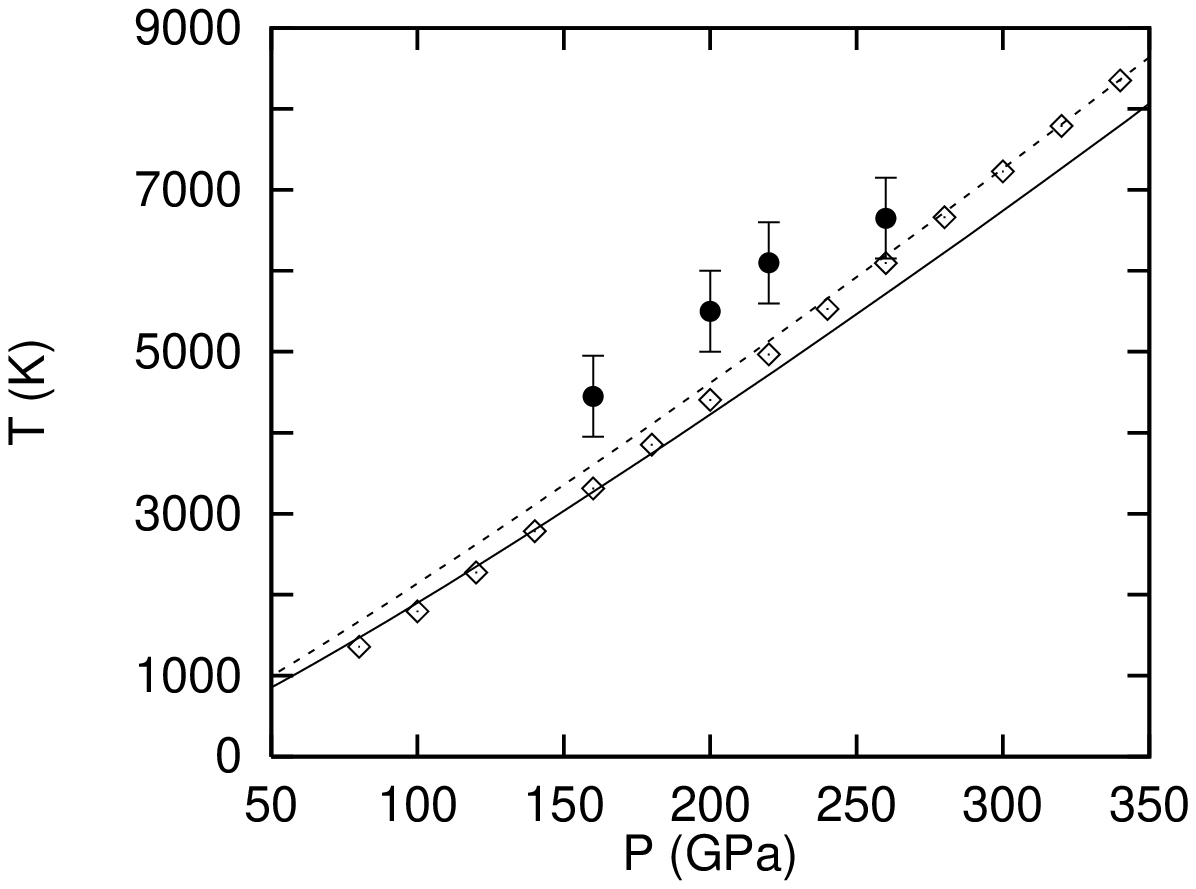,height=3in}}
\clearpage
\centerline{FIGURE 9}
\centerline{\psfig{figure=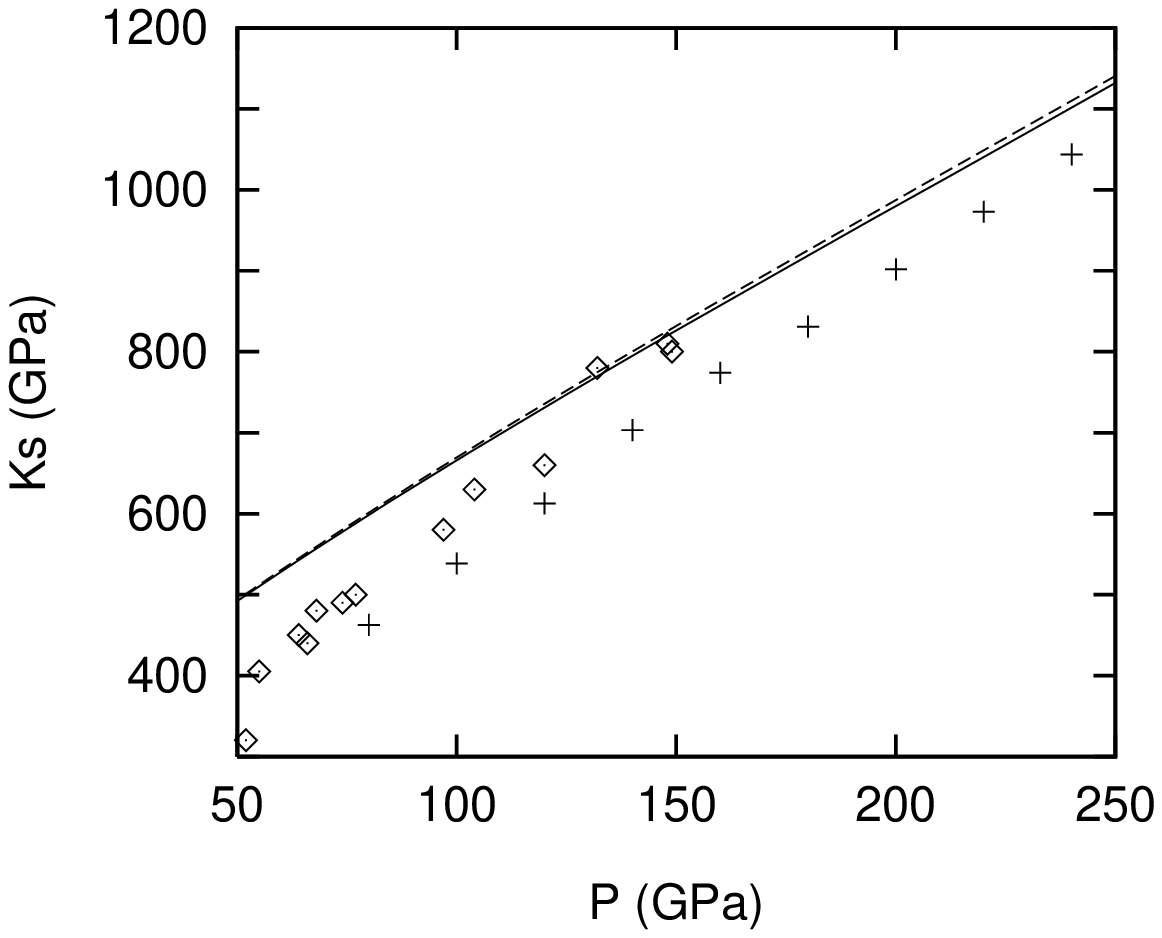,height=3in}}
\clearpage
\centerline{FIGURE 10}
\centerline{\psfig{figure=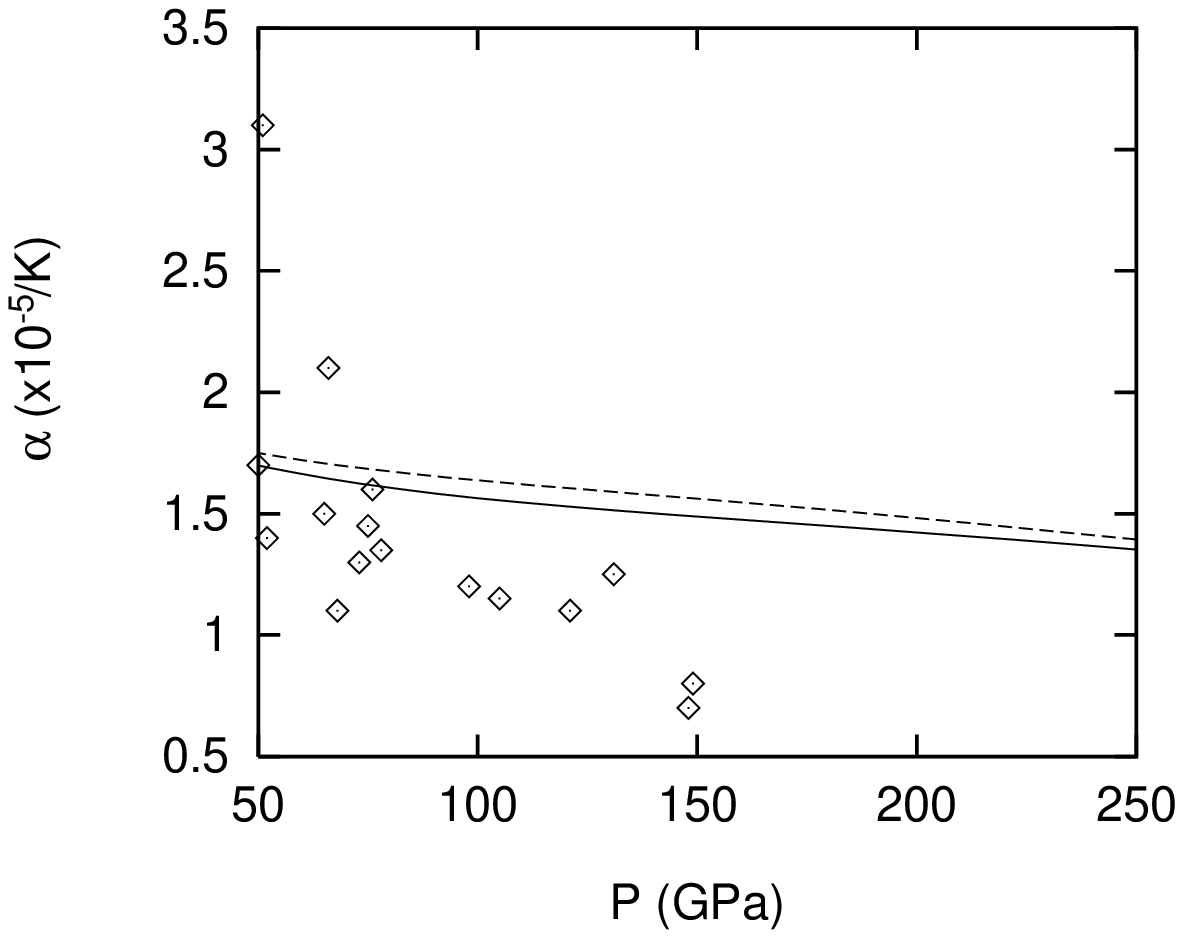,height=3in}}
\clearpage
\centerline{FIGURE 11}
\centerline{\psfig{figure=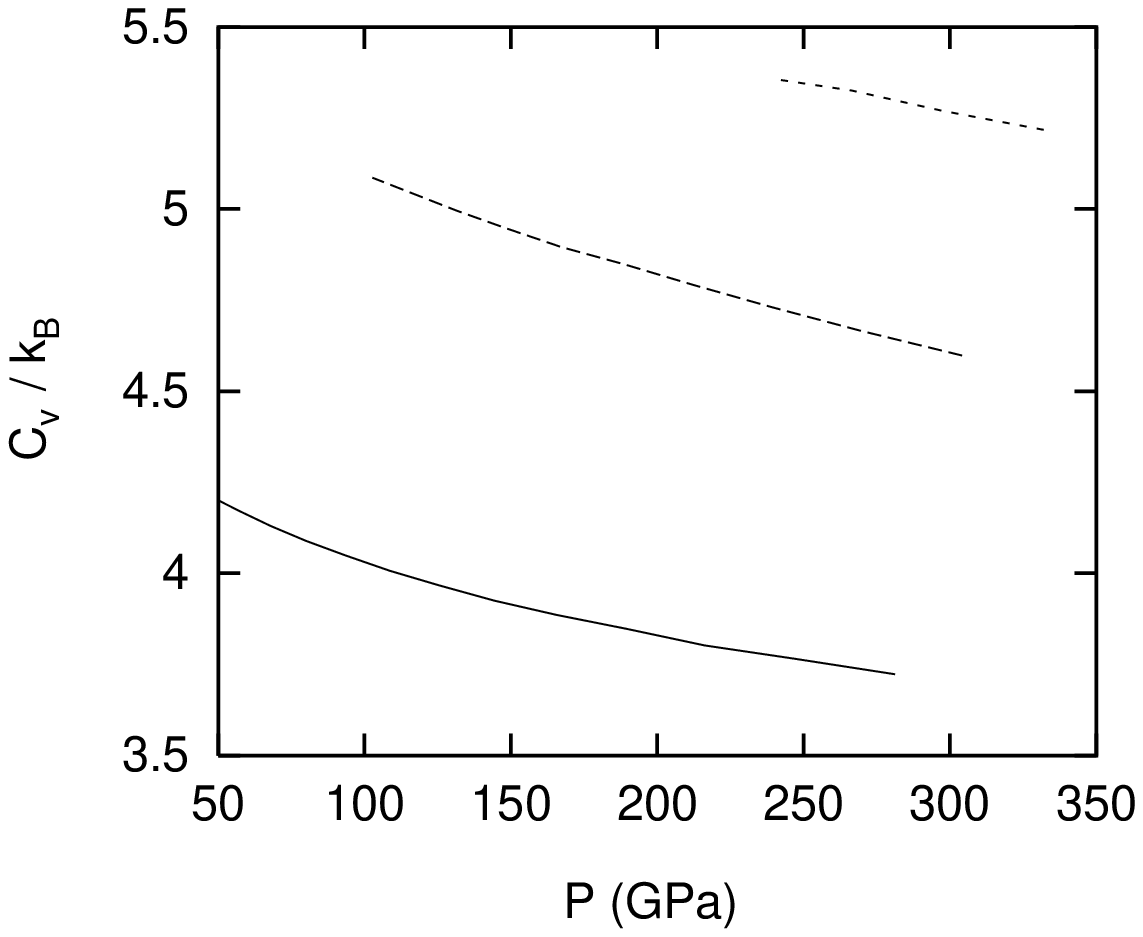,height=3in}}
\clearpage
\centerline{FIGURE 12}
\centerline{\psfig{figure=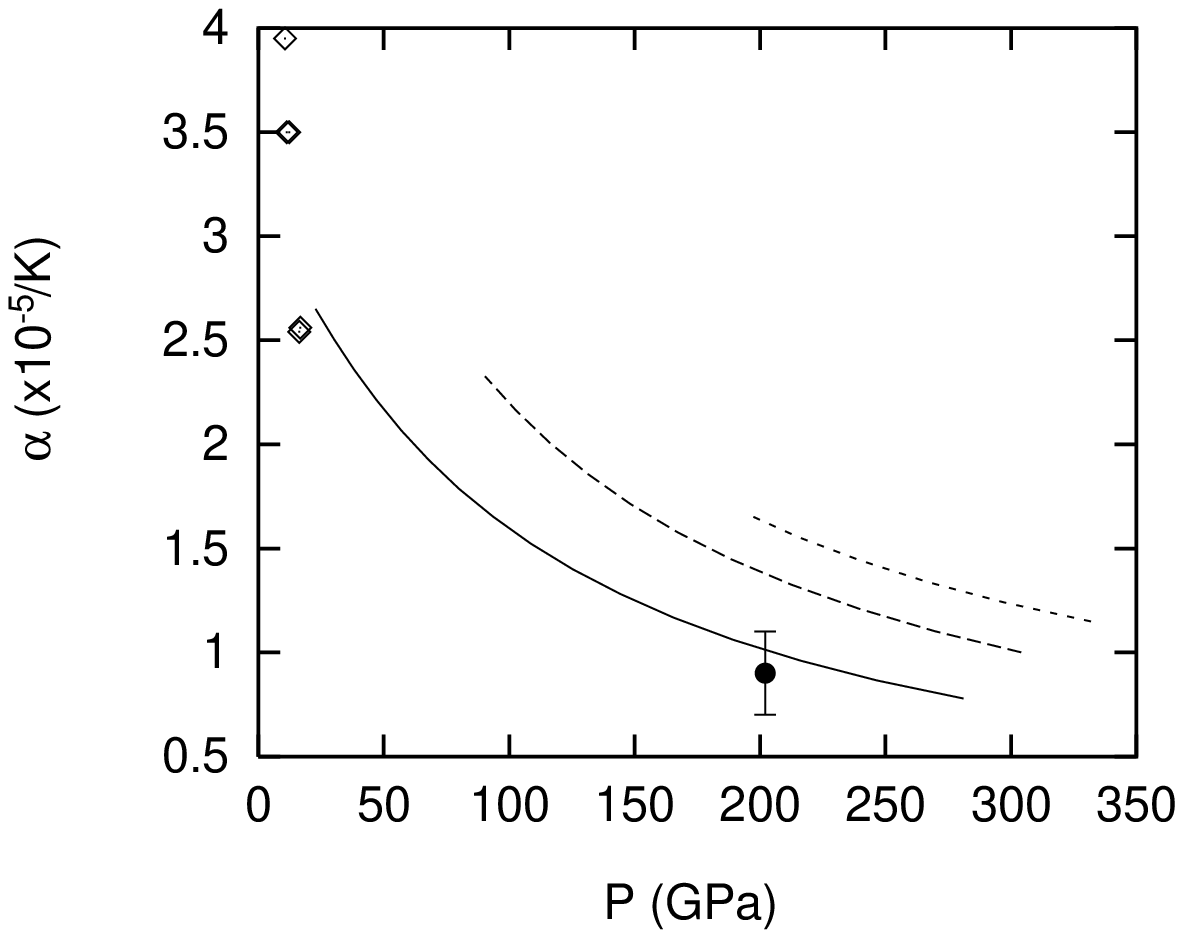,height=3in}}
\clearpage
\centerline{FIGURE 13}
\centerline{\psfig{figure=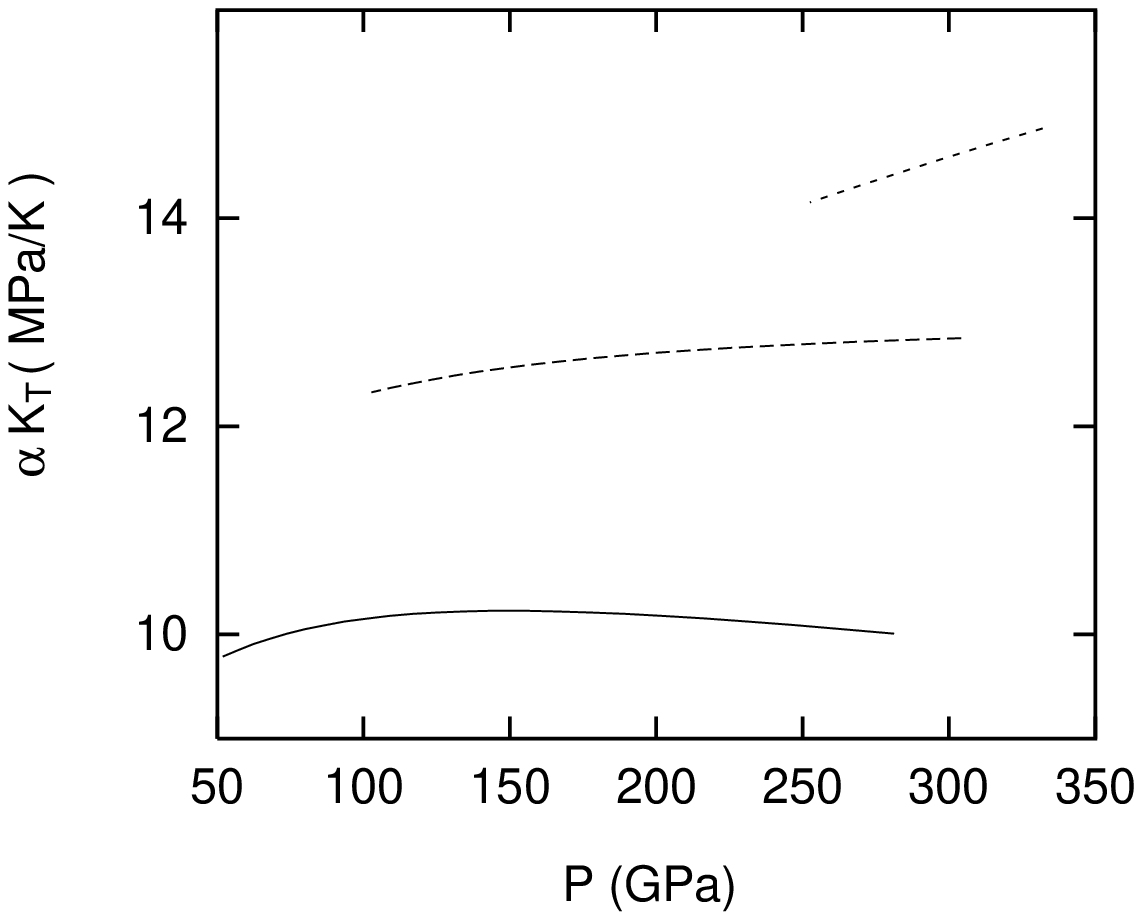,height=3in}}
\clearpage
\centerline{FIGURE 14}
\centerline{\psfig{figure=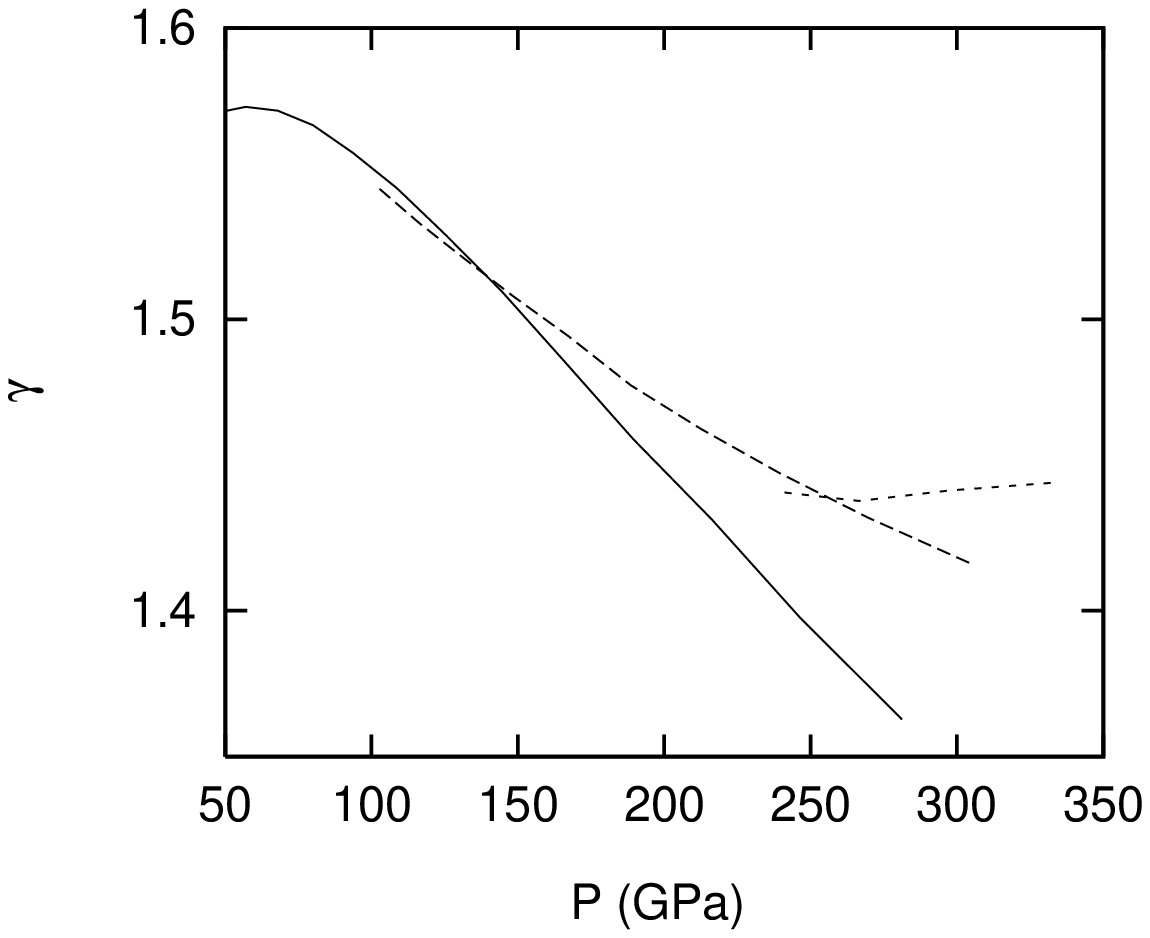,height=3in}}


\begin{thebibliography}{99}

\bibitem{stixrude98}
L. Stixrude, R. E. Cohen and R. J. Hemley, Rev. Mineralogy {\bf 37},
639 (1998).

\bibitem{sugino95}
O. Sugino and R. Car, Phys. Rev. Lett. {\bf 74}, 1823 (1995).

\bibitem{dewijs98a}
G. A. de Wijs, G. Kresse and M. J. Gillan, Phys. Rev. B {\bf 57},
8223 (1998).

\bibitem{alfe99a}
D. Alf\`{e}, G. A. de Wijs, G. Kresse and M. J. Gillan,
Int. J. Quant. Chem., in press.

\bibitem{poirier91}
J.-P. Poirier, {\em Introduction to the Physics of the Earth's
Interior}, Cambridge University Press, Cambridge (1991).

\bibitem{poirier94}
J.-P. Poirier, Phys. Earth Planet. Inter. {\bf 85}, 319 (1994).

\bibitem{masters90}
T. G. Masters and P. M. Shearer, J. Geophys. Res. {\bf 95}, 21691 (1990).

\bibitem{stixrude97}
L. Stixrude, E. Wasserman and R. E. Cohen, J. Geophys. Res.
{\bf 102}, 24729 (1997).

\bibitem{mao90}
H. K. Mao, Y. Wu, L. C. Chen, J. F. Shu and A. P. Jephcoat,
J. Geophys. Res. {\bf 95}, 21737 (1990).

\bibitem{boehler90}
R. Boehler, N. von Bargen and A. Chopelas, J. Geophys. Res.
{\bf 95}, 731 (1990). 

\bibitem{boehler93}
R. Boehler, Nature {\bf 363}, 534 (1993).

\bibitem{saxena93}
S. K. Saxena, G. Shen and P, Lazor, Science {\bf 260}, 1312 (1993).

\bibitem{saxena94}
S. K. Saxena, G. Shen and P. Lazor, Science {\bf 264}, 405 (1994).

\bibitem{saxena95}
S. K. Saxena, L. S. Dubrovinsky, P. H\"{a}ggkvist, Y. Cerenius, G. Shen
and H. K. Mao, Science {\bf 269}, 1703 (1995).

\bibitem{saxena96}
S. K. Saxena, L. S. Dubrovinsky and P. H\"{a}ggkvist,
Geophys. Res. Lett. {\bf 23}, 2441 (1996).

\bibitem{jephcoat96}
A. P. Jephcoat and S. P. Besedin, Phil. Trans. Roy. Soc. Series A
{\bf 354}, 1333 (1996).

\bibitem{andrault97}
D. Andrault, G. Fiquet, M. Kunz, F. Visocekas and D. H\"{a}usermann,
Science {\bf 278}, 831 (1997). 

\bibitem{shen98}
G. Shen, H. Mao, R. J. Hemley, T. S. Duffy and M. L. Rivers,
Geophys. Res. Lett. {\bf 25}, 373 (1998).

\bibitem{vocadlo99}
L. Vo\v{c}adlo, J. Brodholt, D. Alf\`{e}, M. J. Gillan and G. D. Price,
Phys. Earth Planet. Inter., in press.

\bibitem{jeanloz79}
R. Jeanloz, J. Geophys. Res. {\bf 84}, 6059 (1979).

\bibitem{brown86}
J. M. Brown and R. G. McQueen, J. Geophys. Res. {\bf 91},
7485 (1986).

\bibitem{yoo93}
C. S. Yoo, N. C. Holmes, M. Ross, D. J. Webb and C. Pike,
Phys. Rev. Lett. {\bf 70}, 3931 (1993).

\bibitem{matsui97}
M. Matsui and O. L. Anderson, Phys. Earth Planet. Inter. {\bf 103},
55 (1997).

\bibitem{belonoshko97}
A. B. Belonoshko and R. Ahuja, Phys. Earth Planet. Inter. {\bf 102},
171 (1997).

\bibitem{generaldft}
P. Hohenberg and W. Kohn, Phys. Rev. {\bf 136}, B864 (1964);
W. Kohn and L. Sham, Phys. Rev. A1133 (1965); R. O. Jones and O. Gunnarsson,
Rev. Mod. Phys. {\bf 61}, 689 (1989); M. J. Gillan, Contemp. Phys.
{\bf 38}, 115 (1997).

\bibitem{wang85}
C. S. Wang, B. M. Klein and H. Krakauer, Phys. Rev. Lett.
{\bf 54}, 1852 (1985).

\bibitem{stixrude94}
L. Stixrude, R. E. Cohen and D. J. Singh, Phys. Rev. B {\bf 50}, 6442 (1994).

\bibitem{soderlind96}
P. S\"{o}derlind, J. A. Moriarty and J. M. Willis, Phys. Rev. B
{\bf 53}, 14063 (1996).

\bibitem{vocadlo97}
L. Vo\v{c}adlo, G. A. de Wijs, G. Kresse, M. J. Gillan and G. D. Price,
Faraday Disc. {\bf 106}, 205 (1997).

\bibitem{dewijs98b}
G. A. de Wijs, G. Kresse, L. Vo\v{c}adlo, D. Dobson, D. Alf\`{e},
M. J. Gillan and G. D. Price, Nature {\bf 392}, 805 (1998).

\bibitem{alfe99b}
D. Alf\`{e}, G. Kresse and M. J. Gillan, Phys. Rev. B, submitted.

\bibitem{alfe98a}
D. Alf\`{e} and M. J. Gillan, Phys. Rev. B {\bf 58}, 8248 (1998).

\bibitem{alfe98b}
D. Alf\`{e} and M. J. Gillan, Phys. Rev. Lett. {\bf 81}, 5161 (1998).

\bibitem{alfe99c}
D. Alf\`{e}, G. D. Price and M. J. Gillan, Phys. Earth Planet. Inter.
{\bf 110}, 191 (1999).

\bibitem{wasserman96}
E. Wasserman, L. Stixrude and R. E. Cohen, Phys. Rev. B {\bf 53},
8296 (1996).

\bibitem{holt70}
A. C. Holt and M. Ross, Phys. Rev. B {\bf 1}, 2700 (1970).

\bibitem{wang91}
Y. Wang and J. Perdew, Phys. Rev. B {\bf 44}, 13298 (1991).

\bibitem{perdew92}
J. P. Perdew, J. A. Chevary, S. H. Vosko, K. A. Jackson,
M. R. Pederson, D. J. Singh and C. Fiolhais, Phys. Rev. B
{\bf 46}, 6671 (1992).

\bibitem{blochl94}
P. E. Bl\"{o}chl, Phys. Rev. B {\bf 50}, 17953 (1994).

\bibitem{kresse99}
G. Kresse and D. Joubert, Phys. Rev. B {\bf 59}, 1758 (1999).

\bibitem{wei85}
S. Wei and H. Krakauer, Phys. Rev. Lett.
{\bf 55}, 1200 (1985).

\bibitem{alfe99d}
A preliminary report of our calculations on the melting curve
of Fe will appear in D. Alf\`{e}, G. D. Price and 
M. J. Gillan, Nature, in press.

\bibitem{vanderbilt90}
D. Vanderbilt, Phys. Rev. B {\bf 41}, 7892 (1990).

\bibitem{mermin65}
N. D. Mermin, Phys. Rev. {\bf 137}, A1441 (1965).

\bibitem{gillan89}
M. J. Gillan, J. Phys. Condens. Matter {\bf 1}, 689 (1989).

\bibitem{wentzcovitch92}
R. M. Wentzcovitch, J. L. Martins and P. B. Allen,
Phys. Rev. B {\bf 45}, 11372 (1992).

\bibitem{kresse96a}
G. Kresse and J. Furthm\"{u}ller, Phys. Rev. B {\bf 54},
11169 (1996).

\bibitem{kresse96b}
G. Kresse and J. Furthm\"{u}ller, Comput. Mater. Sci. {\bf 6},
15 (1996).

\bibitem{monkhorst76}
H. J. Monkhorst and J. D. Pack, Phys. Rev. B {\bf 13}, 5188 (1976).

\bibitem{kresse95}
G. Kresse,
J. Furthm\"{u}ller and J. Hafner, Europhys. Lett. {\bf 32},
729 (1995).

\bibitem{frenkel96}
D. Frenkel and B. Smit, {\em Understanding Molecular Simulation},
Academic Press, San Diego (1996).

\bibitem{andersen80}
H. C. Andersen, J. Chem. Phys. {\bf 72}, 2384 (1980).

\bibitem{boness90}
D. A. Boness 
and J. M. Brown, J. Geophys. Res. {\bf 95}, 21721 (1990).

\bibitem{ashcroft76}
N. W. Ashcroft and N. D. Mermin, {\em Solid State Physics}, ch.~2,
Holt, Rinehart and Winston, New York (1976).

\bibitem{anderson95}
It is often assumed that the thermodynamic Gr\"{u}neisen
parameter $\gamma$ is independent of temperature and depends
on volume as $(V/V_0 )^q$, where $V_0$ is a reference volume and $q$
is a positive exponent roughly equal to unity; see e.g.
O. L. Anderson, {\em Equations of State of Solids for Geophysics
and Ceramic Science}, Oxford Monographs on Geology and Geophysics No. 31,
Oxford University Press (1995).

\bibitem{vosko80}
S. H. Vosko, L. Wilk and M. Nusair, Can. J. Phys. {\bf 58},
1200 (1980).

\bibitem{rankinehugoniot}
See e.g. ch.~4 of Ref.~5.

\bibitem{duffy93}
T. S. Duffy and T. J. Ahrens, Geophys. Res. Lett. {\bf 20},
1103 (1993).

\bibitem{not_measured}
In contrast to Yoo {\em et al.} (Ref.~22), Brown
and McQueen did not measure temperature in their shock
experiments, but estimated it using an empirical thermodynamic
relation.

\bibitem{stacey95}
F. Stacey, Phys. Earth Planet. Inter. {\bf 89}, 219 (1995).


\end{thebibliography}
\end{document}